\documentclass[final,5p,times,sort&compress]{elsarticle}

\usepackage{amssymb}
\usepackage{amsthm}
\usepackage{natbib}

\usepackage{multirow}
\usepackage{multicol}

\journal{Physica E}

\begin{document}

\begin{frontmatter}

\title{On the failure of effective-field theory in predicting a spurious spontaneous ordering and phase transition of Ising nanoparticles, nanoislands, nanotubes and nanowires}

\author[label1]{Jozef Stre\v{c}ka\corref{cor1}}
\cortext[cor1]{Corresponding author}
\ead{jozef.strecka@upjs.sk}
\author[label1]{Katar\'ina Karl'ov\'a}
\author[label1,label2]{Olesia Krupnitska}
\address[label1]{Department of Theoretical Physics and Astrophysics, Faculty of Science,
					       P.~J.~\v{S}af\'arik University, Park Angelinum 9, 040\,01 Ko\v{s}ice, Slovakia}
\address[label2]{Institute for Condensed Matter Physics, National Academy of Sciences of Ukraine, 
                 Svientsitskii Street 1, 790\,11 L'viv, Ukraine}

\begin{abstract}
The present work clarifies a failure of the effective-field theory in predicting a false spontaneous long-range order and phase transition of Ising nanoparticles, nanoislands, nanotubes and nanowires with either zero- or one-dimensional magnetic dimensionality. It is conjectured that the standard formulation of the effective-field theory due to Honmura and Kaneyoshi generally predicts for the Ising spin systems a spurious spontaneous long-range order with nonzero critical temperature regardless of their magnetic dimensionality whenever at least one Ising spin has coordination number greater than two. The failure of the effective-field theory is exemplified on a few paradigmatic exactly solved examples of zero- and one-dimensional Ising nanosystems: star, cube, decorated hexagon, star of David, branched chain, sawtooth chain, two-leg and hexagonal ladders. The presented exact solutions illustrate eligibility of a few rigorous analytical methods for exact treatment of the Ising nanosystems: exact enumeration, graph-theoretical approach, transfer-matrix method and decoration-iteration transformation. The paper also provides a substantial survey of the scientific literature, in which the effective-field theory led to a false prediction of the spontaneous long-range ordering and phase transition.     
\end{abstract}



\end{frontmatter}


\section{Introduction}
\label{intro}

The effective-field theory (EFT) based on the differential operator technique has been suggested by Honmura and Kaneyoshi \cite{honm78,honm79} as a relatively simple and versatile approximate calculation method, which however provides for the Ising model quite superior results with respect to the standard mean-field theory \cite{kane93,stre15}. The conceptual simplicity of the EFT has allowed its application to numerous Ising models of very different character (see Refs. \cite{kane93,stre15} and references cited therein). In contrast to the mean-field method the EFT correctly reproduces also zero critical temperature of the simple spin-1/2 Ising chain \cite{kane93,stre15}, which could imply that it provides a much more reliable method for low-dimensional Ising models as well. However, it will be demonstrated hereafter that the EFT suffers from serious deficiency in that it is incapable of predicting absence of a spontaneous long-range order and phase transition in zero- or one-dimensional Ising spin systems. 

It is beyond the scope of the present paper to review here all calculation details of the EFT, so let us merely quote a few basic steps of this computational method. The readers interested in further details are referred to extensive review articles \cite{kane93,stre15}. The starting point of the EFT is the exact Callen-Suzuki spin identity \cite{call63,suzu65,balc02}:
\begin{eqnarray}
\langle S_i \rangle  = \left\langle \frac{\sum_{S_i} S_i \exp(-\beta H_i)}
{\sum_{S_i} \exp(-\beta H_i)}\right\rangle, \qquad \left(\beta = \frac{1}{k_B T}\right)
\label{csi}
\end{eqnarray}
which allows a simple calculation of the magnetization by considering the local Hamiltonian $H_i$ involving all interaction terms of the central spin $S_i$. Taking advantage of the differential operator technique $\exp(\alpha \nabla_x) f (x) = f (x + \alpha)$ and the exact van der Waerden spin identity \cite{waer41} $\exp(\alpha S_j) = \cosh (\alpha) + S_j \sinh(\alpha)$ one consequently gets an exact expression for the magnetization, which generally depends on higher-order spin correlators $\langle S_i S_j \cdots S_k \rangle$. Within the framework of the standard formulation of the EFT \cite{kane93} one further introduces a decoupling approximation for higher-order spin correlators $\langle S_i S_j \cdots S_k \rangle \cong \langle S_i \rangle \langle S_j \rangle \cdots \langle S_k \rangle$, which represents the only approximate step of this calculation procedure splitting certain spin-spin correlations. Besides a straightforward calculation of the magnetization, the EFT can be easily adapted for calculation of the statistical mean values for the product of a central spin $S_i$ with any function of all other spins $f(\{S_j \}_{j \neq i})$ according to the generalized Callen-Suzuki spin identity \cite{call63,suzu65,balc02}:
\begin{eqnarray}
\langle S_i f (\{S_j \}_{j \neq i})\rangle  = \left\langle f (\{S_j \}_{j \neq i}) 
\frac{\sum_{S_i} S_i \exp(-\beta H_i)}{\sum_{S_i} \exp(-\beta H_i)}\right\rangle.
\label{csig}
\end{eqnarray}
This procedure can be for instance employed for the calculation of the pair spin-spin correlator, which successively enables evaluation of basic thermodynamic quantities such as the internal energy and specific heat. 

In the present article we will adapt the standard formulation of the EFT due to Honmura and Kaneyoshi \cite{kane93} in order to calculate the magnetization, the pair spin correlator, the internal energy and the specific heat of a few selected Ising nanosystems with either zero- or one-dimensional magnetic dimension. More specifically, we will examine magnetic and thermodynamic properties of the spin-1/2 Ising star, cube, decorated hexagon, star of David, two-leg and hexagonal ladders. To illustrate a failure of the EFT in reproducing absence of the spontaneous long-range order and phase transition we will also adapt a few rigorous calculation methods, which provide for these low-dimensional Ising spin systems exact results being completely free of any uncontrollable approximation. 

The organization of this paper is as follows. In Sections~\ref{sec:2} and \ref{sec:3} we will clearly demonstrate a failure of the EFT by considering a few paradigmatic examples of zero- and one-dimensional spin-1/2 Ising nanosystems through a comparison with the respective exact results. Section~\ref{sec:4} involves the general arguments about non-existence of phase transition in zero- and one-dimensional spin-1/2 Ising systems besides a comprehensive review on the previously published literature, where the application of EFT caused a false prediction of the spontaneous long-range order and the associated critical behavior. A brief summary of the most important findings is presented in Section~\ref{sec:5}.

\section{Zero-dimensional Ising nanosystems}
\label{sec:2}

In this section we will investigate a few paradigmatic examples of the Ising nanosystems, each of which will be further treated within the framework of the EFT and some exact analytical method. In particular, we will consider the spin-1/2 Ising star, cube, decorated hexagonal nanoparticle and star of David whose magnetic dimension is zero. The calculation procedure based on the EFT will be exemplified in detail on the particular example of the spin-1/2 Ising star, while only a few basic steps of this method will be presented for the remaining spin clusters. 

\begin{figure}[t]
\begin{center}
\includegraphics[width=0.8\columnwidth]{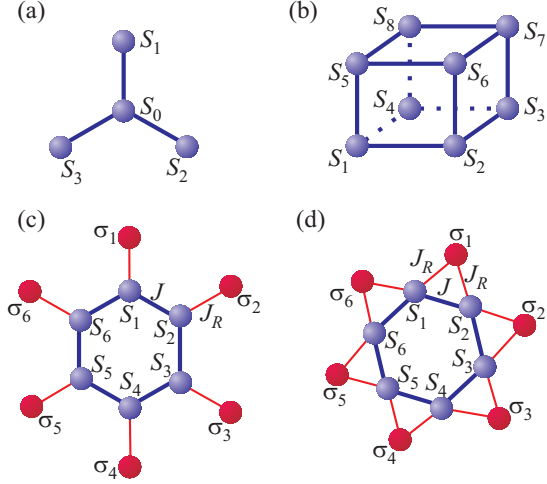}
\end{center}
\vspace*{-0.5cm}
\caption{A few typical examples of zero-dimensional Ising spin nanosystems: (a) star; (b) cube; (c) decorated hexagonal nanoparticle; (d) star of David.}
\label{fig0d}
\end{figure}

\subsection{Star}
\label{sec:star}

Let us consider first the spin-1/2 Ising star as the simplest nanosystem, for which the EFT fails to reproduce absence of the spontaneous long-range order. The spin-1/2 Ising star refers to a four-spin cluster schematically shown in Fig. \ref{fig0d}(a) and mathematically given by the Hamiltonian:
\begin{eqnarray}
H = - J S_0 (S_1 + S_2 + S_3),
\label{eq:hams}
\end{eqnarray}  
where $S_0 = \pm 1$ is the Ising spin attached to a central site '0' with the coordination number three and $S_j = \pm 1$ ($j=1,2,3$) are three Ising spins attached to boundary sites '1-3' with the coordination number one. Although the partition function and the whole thermodynamics of the spin-1/2 Ising star can be exactly calculated by performing a simple summation over all available spin configurations of the four underlying Ising spins, we will at first illustrate the mathematical treatment taking advantages of the EFT.

\subsubsection{Effective-field theory}   
\begin{figure}[t]
\begin{center}
\includegraphics[width=0.9\columnwidth]{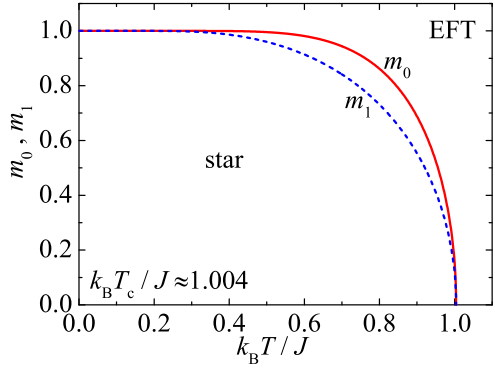}
\end{center}
\vspace*{-0.5cm}
\caption{Temperature variations of the spontaneous magnetization of the central and boundary spins calculated for the spin-1/2 Ising star within the EFT according to Eq. (\ref{eq:sm}).}
\label{figsm}
\end{figure}

The mean values for the local magnetizations of the central and boundary spins of the spin-1/2 Ising star can be calculated from the exact Callen-Suzuki spin identities \cite{call63,suzu65,balc02}:
\begin{eqnarray}
m_0 \!\!\!&\equiv&\!\!\! \langle S_0 \rangle = \langle \tanh[\beta J (S_1 + S_2 + S_3)] \rangle, \nonumber \\
m_1 \!\!\!&\equiv&\!\!\! \langle S_j \rangle = \langle \tanh(\beta J S_0) \rangle, \qquad (j = 1,2,3). 
\label{eq:smag}
\end{eqnarray}  
Using the differential operator technique $\exp(\alpha \nabla_x) f (x) = f (x + \alpha)$ the couple of equations (\ref{eq:smag}) can be rewritten as follows:
\begin{eqnarray}
m_0 \!\!\!&=&\!\!\! \langle \exp[(S_1 + S_2 + S_3) \nabla_x]  \rangle \tanh(\beta J x), \nonumber \\
m_1 \!\!\!&=&\!\!\! \langle \exp(S_0 \nabla_x) \rangle \tanh(\beta J x). 
\label{eq:smdo}
\end{eqnarray}
The subsequent application of the exact van der Waerden spin identity \cite{waer41} $\exp(\alpha S_j) = \cosh (\alpha) + S_j \sinh(\alpha)$ together with the differential operator technique leads to the following equations:
\begin{eqnarray}
m_0 \!\!\!&=&\!\!\! \langle S_1 + S_2 + S_3 \rangle K_1 + \langle S_1 S_2 S_3 \rangle K_2, \nonumber \\
m_1 \!\!\!&=&\!\!\! \langle S_0 \rangle K_0, 
\label{eq:smvw}
\end{eqnarray} 
which include the coefficients $K_0$, $K_1$ and $K_2$ defined as follows:
\begin{eqnarray}
K_0 \!\!\!&=&\!\!\! \tanh(\beta J), \nonumber \\
K_1 \!\!\!&=&\!\!\! \frac{1}{4} \left[\tanh (3\beta J) + \tanh (\beta J) \right],  \nonumber \\
K_2 \!\!\!&=&\!\!\! \frac{1}{4} \left[\tanh (3\beta J) - 3 \tanh (\beta J) \right]. 
\label{eq:sc}
\end{eqnarray} 
It is noteworthy that the set of equations (\ref{eq:smvw}) derived for the local magnetizations is still exact. Within the standard formulation of the EFT one further takes advantage of Honmura-Kaneyoshi (HK) decoupling scheme for all higher-order spin correlators $\langle S_i S_j \cdots S_k \rangle \cong \langle S_i \rangle \langle S_j \rangle \cdots \langle S_k \rangle$ being equivalent to Zernike approximation \cite{kane93}. Consequently, one arrives at two interconnected equations for the local magnetizations:
\begin{eqnarray}
m_0 \!\!\!&=&\!\!\! 3 m_1 K_1 + m_1^3 K_2, \nonumber \\
m_1 \!\!\!&=&\!\!\! m_0 K_0.
\label{eq:smi}
\end{eqnarray} 
The subsequent elimination either of the local magnetization $m_0$ or $m_1$ from the couple of equations (\ref{eq:smi}) provides the following results for the local magnetizations of the spin-1/2 Ising star:
\begin{eqnarray}
m_0 \!\!\!&=&\!\!\! \sqrt{\frac{1 - 3 K_0 K_1}{K_0^3 K_2}}, \nonumber \\
m_1 \!\!\!&=&\!\!\! \sqrt{\frac{1 - 3 K_0 K_1}{K_0 K_2}}.
\label{eq:sm}
\end{eqnarray} 
Temperature dependences of the spontaneous magnetizations $m_0$ and $m_1$ of the central and boundary spins of the spin-1/2 Ising star calculated according to Eq.~(\ref{eq:sm}) are presented in Fig.~\ref{figsm}. The local magnetization for the central spin $m_0$ is generally more resistant against thermal fluctuations than the local magnetization for the boundary spins $m_1$ because of a higher coordination number of the central spin with respect to the boundary ones.  
In agreement with the formulas (\ref{eq:sm}) both local spontaneous magnetizations however tend to zero at the critical temperature $k_B T_c/J \approx 1.004$, which can be calculated from the condition $3 K_0 K_1 = 1$ or equivalently:
\begin{eqnarray}
3 \tanh (\beta_c J) [\tanh (3 \beta_c J) + \tanh (\beta_c J)] = 4.
\label{eq:tc}
\end{eqnarray}  

The generalized Callen-Suzuki identity \cite{call63,suzu65,balc02} can be also employed for a calculation of the pair correlator $\langle S_0 S_1 \rangle$, 
\begin{eqnarray}
\langle S_0 S_1 \rangle = \langle S_1 \tanh[\beta J (S_1 + S_2 + S_3)] \rangle, 
\label{eq:scor}
\end{eqnarray} 
which can be straightforwardly used in order to calculate the internal energy $U = - 3 J \langle S_0 S_1 \rangle$ and other important thermodynamic quantities as for instance the specific heat $C = \partial U/\partial T$. By using the differential operator technique, the exact van der Waerden identity \cite{waer41} and the HK decoupling \cite{kane93} one arrives at the following final result for the pair correlator:
\begin{eqnarray}
\langle S_0 S_1 \rangle = K_1 + m_1^2 (2K_1 + K_2). 
\label{eq:scorf}
\end{eqnarray}

\subsubsection{Exact calculations}

Magnetic and thermodynamic properties of the spin-1/2 Ising star given by the Hamiltonian (\ref{eq:hams}) can be easily  calculated by exact means. The local magnetizations of the central and boundary spins of the spin-1/2 Ising star can be computed as standard canonical ensemble averages:
\begin{eqnarray}
m_j \equiv \langle S_j \rangle = \frac{1}{Z} \sum_{ \{S\} } S_j \exp(-\beta H), \quad (j = 0,1,2,3)
\label{sme}
\end{eqnarray}
where $Z = \sum_{ \{S\} } \exp(-\beta H)$ denotes the canonical partition function and the summation $\sum_{ \{S\} }$ runs over all available configurations of all four spins. It is convenient to calculate the canonical partition function by a consecutive summation over all available spin states of the four spins:
\begin{eqnarray}
Z \!\!\!&=&\!\!\! \sum_{ \{S\} } \exp[\beta J S_0 (S_1 + S_2 + S_3)] \nonumber \\
  \!\!\!&=&\!\!\! \sum_{S_1} \sum_{S_2} \sum_{S_3} 2 \cosh[\beta J (S_1 + S_2 + S_3)] \nonumber \\
  \!\!\!&=&\!\!\! 4 \cosh(3 \beta J) + 12 \cosh(\beta J).	
\label{smez}
\end{eqnarray}
One can easily convince oneself that the expression entering into the nominator of Eq. (\ref{sme}) identically equals zero after performing summation over all possible spin states:
\begin{eqnarray}
\sum_{ \{S\} } \!\!\!\!\!\!&&\!\!\!\!\!\! 
S_0 \exp(-\beta H) = \sum_{ \{S\} } S_0 \exp[\beta J S_0 (S_1 + S_2 + S_3)] \nonumber \\
  \!\!\!&=&\!\!\! \sum_{S_1} \sum_{S_2} \sum_{S_3} 2 \sinh[\beta J (S_1 + S_2 + S_3)] = 0. 
\label{smem}
\end{eqnarray}
Hence, it follows from Eq. (\ref{smem}) that the spontaneous magnetization of the spin-1/2 Ising star equals  zero, whereas the nonzero spontaneous magnetizations presented in Fig. \ref{figsm} according to Eq. (\ref{eq:sm}) are just artifacts of the HK decoupling scheme exploited within the EFT \cite{kane93}.

\begin{figure}[t]
\begin{center}
\includegraphics[width=0.9\columnwidth]{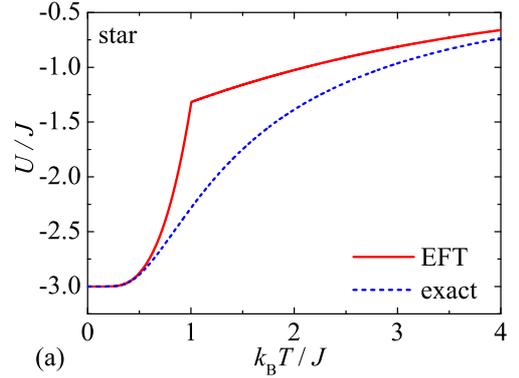}
\includegraphics[width=0.9\columnwidth]{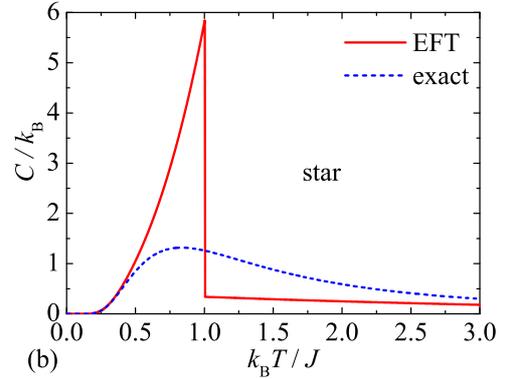}
\end{center}
\vspace*{-0.5cm}
\caption{Temperature variations of the internal energy (panel a) and specific heat (panel b) calculated for the spin-1/2 Ising star within the EFT (solid lines) and exact method (broken lines).}
\label{figsuc}
\end{figure}

To provide independent verification of absence of spontaneous long-range order of the spin-1/2 Ising star one may take advantage of the exact result (\ref{smez}) for the partition function. The internal energy of the spin-1/2 Ising star can be calculated from the partition function (\ref{smez}) according to the relation:
\begin{eqnarray}
U = -\frac{\partial \ln Z}{\partial \beta} 
  = -\frac{12 J}{Z} \left[\sinh(3 \beta J) + \sinh(\beta J) \right],	
\label{smeu}
\end{eqnarray}
while the specific heat can be derived from Eq. (\ref{smeu}) using the basic thermodynamic relation $C = \partial U/\partial T$. For the sake of comparison, the exact results for the internal energy and specific heat of the spin-1/2 Ising star are plotted in Fig. \ref{figsuc} together with the analogous results derived with the help of EFT. It is quite evident that the results derived using the EFT and exact methods coincide in the low- and high-temperature region, while they substantially deviate at moderate temperatures. Moreover, the EFT evidently implies presence of a continuous phase transition associated with a breakdown of the spontaneous long-range order at the critical temperature $k_B T_c/J \approx 1.004$, which manifests itself through a cusp in the temperature dependence of the internal energy and a finite jump in the temperature dependence of the specific heat. Exact results presented in Fig. \ref{figsuc} contrarily indicate smooth continuous temperature variations of the internal energy and specific heat, which apparently preclude presence of a finite-temperature phase transition and spontaneous long-range order. The exact results thus obviously disqualify feasibility of the EFT for the analysis of magnetic and thermodynamic properties of this zero-dimensional Ising spin cluster.

\subsection{Cube}
\label{sec:cube}

Another particular example of nanoscopic spin system is the spin-1/2 Ising cube, which is given by the Hamiltonian:
\begin{eqnarray}
H = \!\!\!&-&\!\!\! J (S_1 S_2 + S_2 S_3 + S_3 S_4 + S_4 S_1) \nonumber \\
    \!\!\!&-&\!\!\! J (S_5 S_6 + S_6 S_7 + S_7 S_8 + S_8 S_5)  \nonumber \\
	  \!\!\!&-&\!\!\! J (S_1 S_5 + S_2 S_6 + S_3 S_7 + S_4 S_8).
\label{eq:hamc}
\end{eqnarray}  
where $S_j = \pm 1$ ($j=1-8$) are the Ising spins located in corners of a simple cube schematically shown in Fig. \ref{fig0d}(b). 

\subsubsection{Effective-field theory} 

\begin{figure}
\includegraphics[width=0.9\columnwidth]{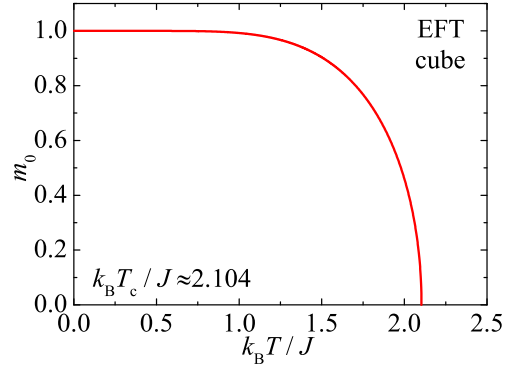}
\vspace*{-0.5cm}
\caption{Temperature variations of the spontaneous magnetization calculated for the spin-1/2 Ising cube 
within the EFT.}
\label{figcm}
\end{figure}

The EFT for the spin-1/2 Ising cube was elaborated by \c{S}arl{\i} in Ref. \cite{sarl15}, so it is sufficient to recall here a few crucial steps of this calculation. The local magnetization of the spin-1/2 Ising cube can be calculated with the help of the exact Callen-Suzuki identity \cite{call63,suzu65,balc02}:
\begin{eqnarray}
m_0 \equiv \langle S_j \rangle = \langle \tanh[\beta J (S_k + S_l + S_m)] \rangle.
\label{eq:cmag}
\end{eqnarray} 
The notation for indices in Eq. (\ref{eq:cmag}) is as follows: the Ising spins $S_k$, $S_l$ and $S_m$ refer to three nearest neighbors with respect to the Ising spin $S_j$. The differential operator technique allows one to rewrite Eq. (\ref{eq:cmag}) into the following form:
\begin{eqnarray}
m_0 = \langle \exp[(S_k + S_l + S_m) \nabla_x]  \rangle \tanh(\beta J x),
\end{eqnarray}
which can be further modified with the help of van der Waerden spin identity \cite{waer41}, the differential operator and the HK decoupling scheme \cite{kane93}:
\begin{eqnarray}
m_0 = 3 m_0 K_1 + m_0^3 K_2.
\label{eq:cmvw}
\end{eqnarray} 
The coefficients $K_1$ and $K_2$ are defined by Eq. (\ref{eq:sc}) and the formula (\ref{eq:cmvw}) can be further put into the following final form:
\begin{eqnarray}
m_0 = \sqrt{\frac{1 - 3 K_1}{K_2}}.
\label{eq:cm}
\end{eqnarray} 
The temperature dependence of the spontaneous magnetization (\ref{eq:cm}) of the spin-1/2 Ising cube is displayed in Fig. \ref{figcm}.
As one can see, the spontaneous magnetization monotonically decreases with increasing temperature until it completely vanishes at the critical temperature $k_B T_c/J \approx 2.104$, which can be calculated from the critical condition $3 K_1 = 1$ or equivalently:
\begin{eqnarray}
3 [\tanh (3 \beta_c J) + \tanh (\beta_c J)] = 4.
\label{eq:ctcs}
\end{eqnarray}  

One may also take advantage of the generalized Callen-Suzuki identity \cite{call63,suzu65,balc02} in order to calculate the pair correlator:
\begin{eqnarray}
\langle S_j S_k \rangle = \langle S_k \tanh[\beta J (S_k + S_l + S_m)] \rangle, 
\label{eq:ccor}
\end{eqnarray} 
which can be used for a calculation of the internal energy $U = - 12 J \langle S_j S_k \rangle$ and the specific heat $C = \partial U/\partial T$. By using the differential operator technique, the exact van der Waerden identity \cite{waer41} and the HK decoupling scheme \cite{kane93} one arrives at the following final result for the pair correlator:
\begin{eqnarray}
\langle S_j S_k \rangle = K_1 + m_0^2 (2K_1 + K_2). 
\label{eq:ccorf}
\end{eqnarray}

\subsubsection{Exact results} 

A full energy spectrum for the spin-1/2 Ising cube can be also calculated by considering a summation over all possible spin configurations, however, in the following we will closely follow a relatively simple graph-theoretical approach comprehensively described in our recent work dealing with the spin-1/2 Ising clusters with the geometric shape of Platonic solids \cite{strec15}. The essence of this rigorous technique lies in finding graphical representations of all possible spin configurations, which can be represented by induced subgraphs whose vertices resemble spins flipped from the fully polarized (ferromagnetic) configuration. One should accordingly start from the ferromagnetic state with the maximal value of the total spin $S_T=8$, which has for the spin-1/2 Ising cube the energy $E_{FM}=-12J$. The energy of other spin configurations obtained from the fully polarized ferromagnetic state by gradual spin flips can be calculated according to the formula:
\begin{eqnarray}
E_{i}=E_{FM}+2J\tilde{d}_i,
\label{eq:energycube}
\end{eqnarray}
where $\tilde{d}_i$ denotes the total number of antiparallel oriented adjacent spin pairs within the $i$th spin configuration. This number is simultaneously equal to the sum of complementary degrees of all vertices of a corresponding induced subgraph schematically shown in Fig. 2 of Ref. \cite{strec15}. It should be pointed out, moreover, that it is sufficient to consider only spin configurations with nonnegative value of the total spin $S_T\geq 0$ due to the time reversal symmetry of the Hamiltonian (\ref{eq:hamc}). After finding degeneracy of all spin configurations, which corresponds to a multiplicity of the corresponding induced subgraph, one may express the partition function as follows:
\begin{eqnarray}
Z=\sum_{\{S_i\}}\exp(-\beta H) = \sum_i g_i \exp(-\beta E_i).
\end{eqnarray}
Here, $g_i$ is the degeneracy factor and $E_i$ is the configurational energy calculated according to Eq. (\ref{eq:energycube}). The configurational energies of the spin-1/2 Ising cube with the ferromagnetic interaction $J>0$ can be retrieved from table 3 of Ref. \cite{strec15} by substituting $J \to -J$ and considering the zero-field case $h=0$. In this way one obtains the exact formula for the partition function of the spin-1/2 Ising cube:
\begin{eqnarray}
Z \!\!\!&=&\!\!\! 64 + 4\cosh(12\beta J) + 32\cosh(6\beta J) \nonumber \\
 \!\!\!&+&\!\!\! 60\cosh(4\beta J) + 96\cosh(2\beta J).
\label{pfc}
\end{eqnarray}
It is worthwhile to remark that the equivalent expression for the partition function of the spin-1/2 Ising cube was obtained by Syozi from the partition function of the spin-1/2 Ising tetrahedron by making use of a dual transformation \cite{syoz55}. It can be easily verified from Eq. (\ref{pfc}) that the spontaneous magnetization of the spin-1/2 Ising cube equals zero: 
\begin{eqnarray}
m_0 \equiv \langle S_j \rangle = \frac{1}{Z} \sum_{ \{S\} } S_j \exp(-\beta H) = 0, 
\label{cme}
\end{eqnarray}
which means the nonzero spontaneous magnetization shown in Fig. \ref{figcm} according to Eq. (\ref{eq:cm}) is again only artifact of the HK decoupling employed within the EFT \cite{kane93}.

\begin{figure}[t]
\includegraphics[width=0.9\columnwidth]{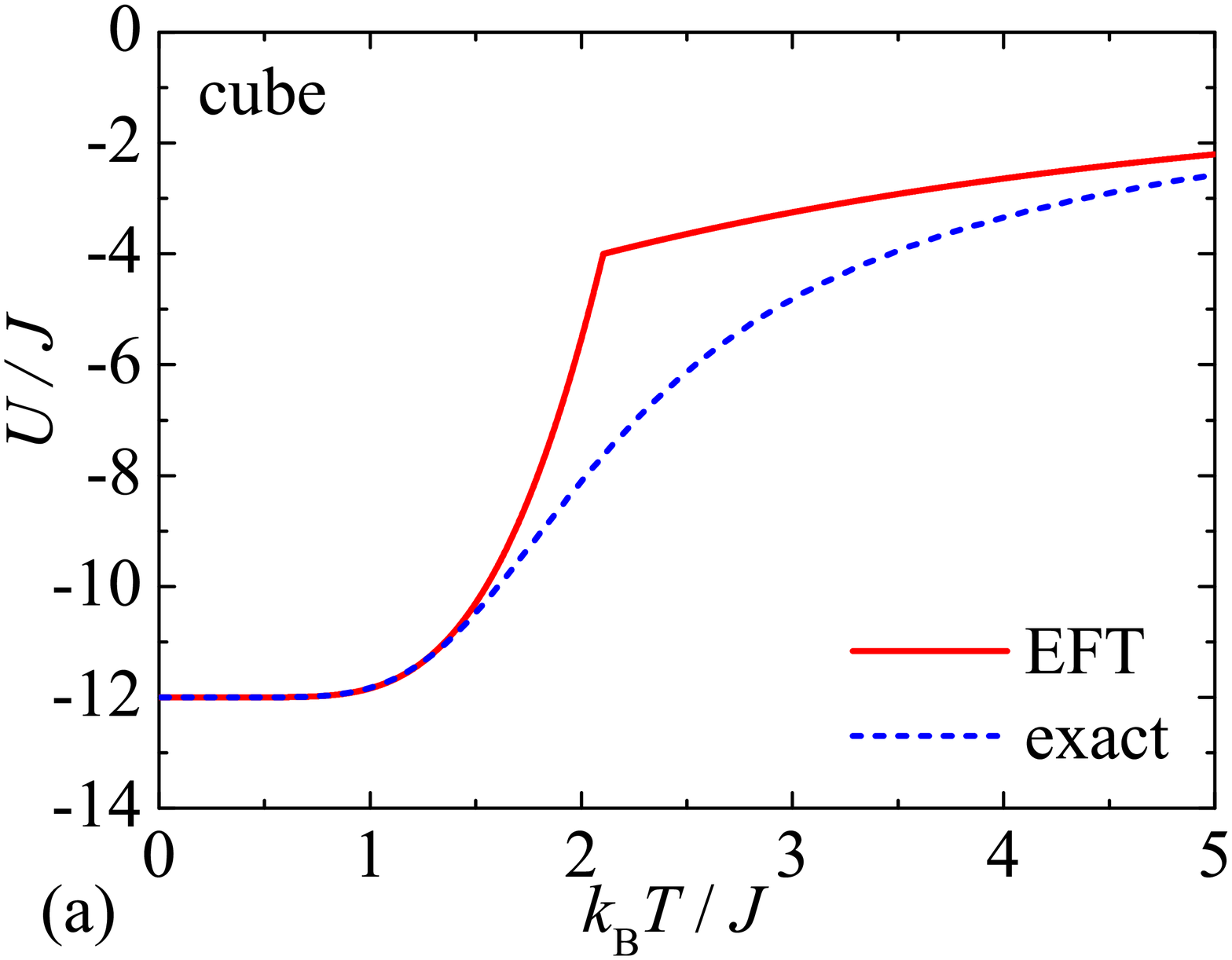}
\includegraphics[width=0.9\columnwidth]{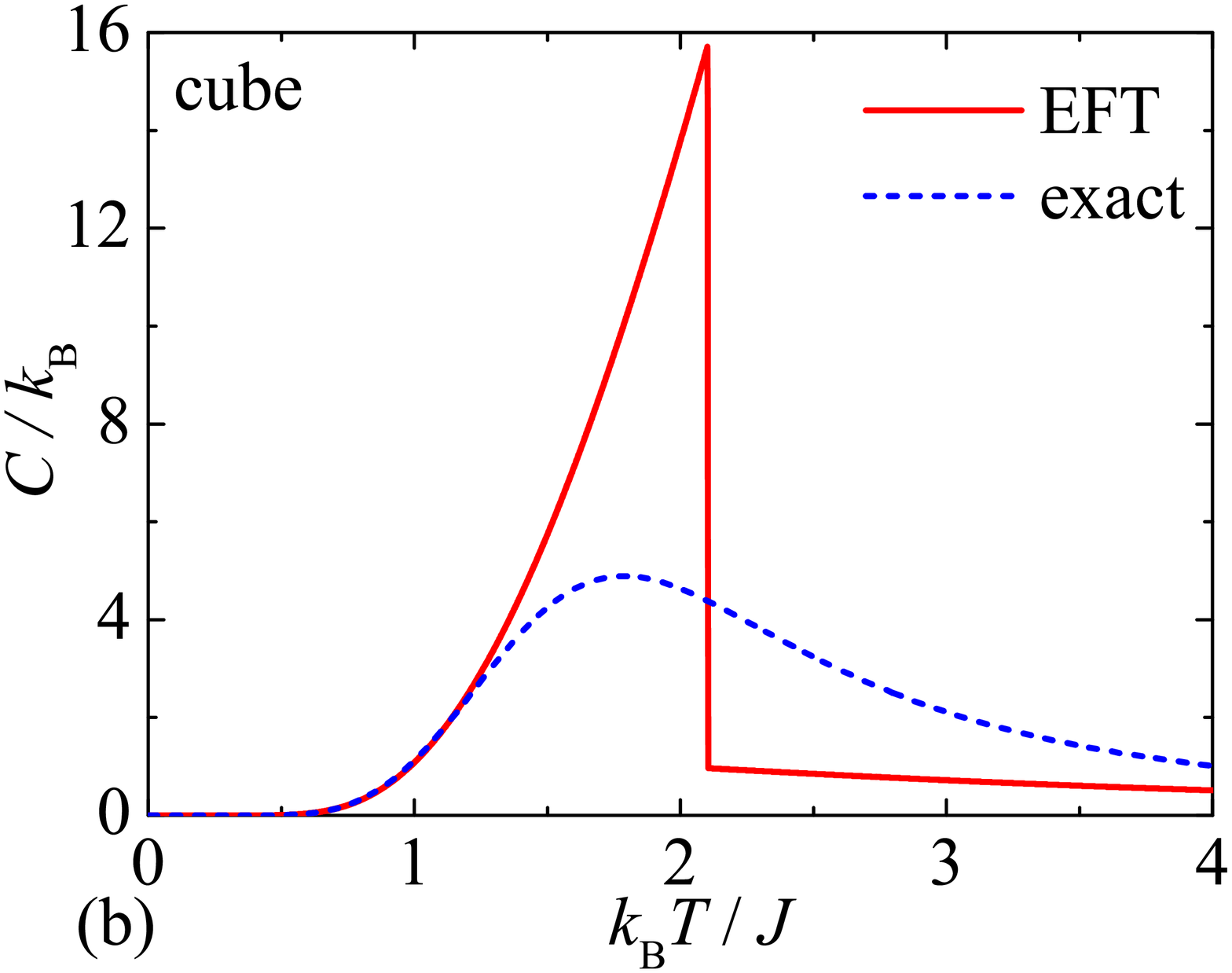}
\vspace*{-0.5cm}
\caption{Temperature variations of the internal energy (panel a) and specific heat (panel b) calculated for the spin-1/2 Ising cube within the EFT (solid lines) and exact method (broken lines).}
\label{figcuc}
\end{figure}

The absence of spontaneous long-range order of the spin-1/2 Ising cube can be also proved from the exact result (\ref{pfc}) for the partition function. The internal energy of the spin-1/2 Ising cube follows from the relation:
\begin{eqnarray}
U = -\frac{\partial \ln Z}{\partial \beta} 
  \!\!\!&=&\!\!\! -\frac{48 J}{Z} \left[\sinh(\beta J) + 4\sinh(6\beta J) \right. \nonumber \\
 \!\!\!&+&\!\!\! \left. 5\sinh(4\beta J) + 4\sinh(2\beta J) \right],	
\label{cmeu}
\end{eqnarray}
while the specific heat can be calculated from Eq. (\ref{cmeu}) using the thermodynamic relation $C = \partial U/\partial T$. The exact results for the internal energy and specific heat of the spin-1/2 Ising cube are compared in Fig. \ref{figcuc} with the analogous results derived with the help of EFT. The results descended from the EFT and exact methods shows the most pronounced discrepancy at moderate temperatures, while they perfectly coincide in the low- and high-temperature regions. According to the EFT, the observed cusp in the temperature dependence of the internal energy and a finite jump in the temperature dependence of the specific heat evidence a continuous phase transition emergent at the critical temperature $k_B T_c/J \approx 2.104$. Contrary to this, the exact results presented in Fig. \ref{figcuc} for the internal energy and specific heat are completely free of any signature of a spontaneous long-range order and a finite-temperature phase transition. It could be thus concluded that the EFT is repeatedly not capable of capturing absence of spontaneous long-range ordering and phase transition of this zero-dimensional Ising spin cluster. 

\subsection{Decorated hexagonal nanoparticle}
\label{sec:hexagon}

We will further consider the spin-1/2 Ising site-decorated nanoparticle given by the Hamiltonian:
\begin{eqnarray}
H = - J \sum_{j=1}^N S_j S_{j+1} - J_R \sum_{j=1}^N S_j \sigma_{j},
\label{eq:hamd}
\end{eqnarray}
which involves the Ising spins $S_j = \pm 1$ and $\sigma_j = \pm 1$ attributed to nodal and decorating sites of the nanoparticle under the periodic boundary condition $S_{N+1} \equiv S_1$. It is noteworthy that the EFT as well as the exact method based on the transfer-matrix approach can be formulated for the spin-1/2 Ising decorated nanoparticle with quite general number of spins $N \geq 3$. It is noteworthy that the specific case with $N=6$, which corresponds to the spin-1/2 Ising decorated hexagon schematically shown in Fig. \ref{fig0d}(c), has been previously studied using the EFT by Kaneyoshi \cite{kane21} and will henceforth serve for illustration. 

\subsubsection{Effective-field theory} 

\begin{figure}[t]
\begin{center}
\includegraphics[width=0.9\columnwidth]{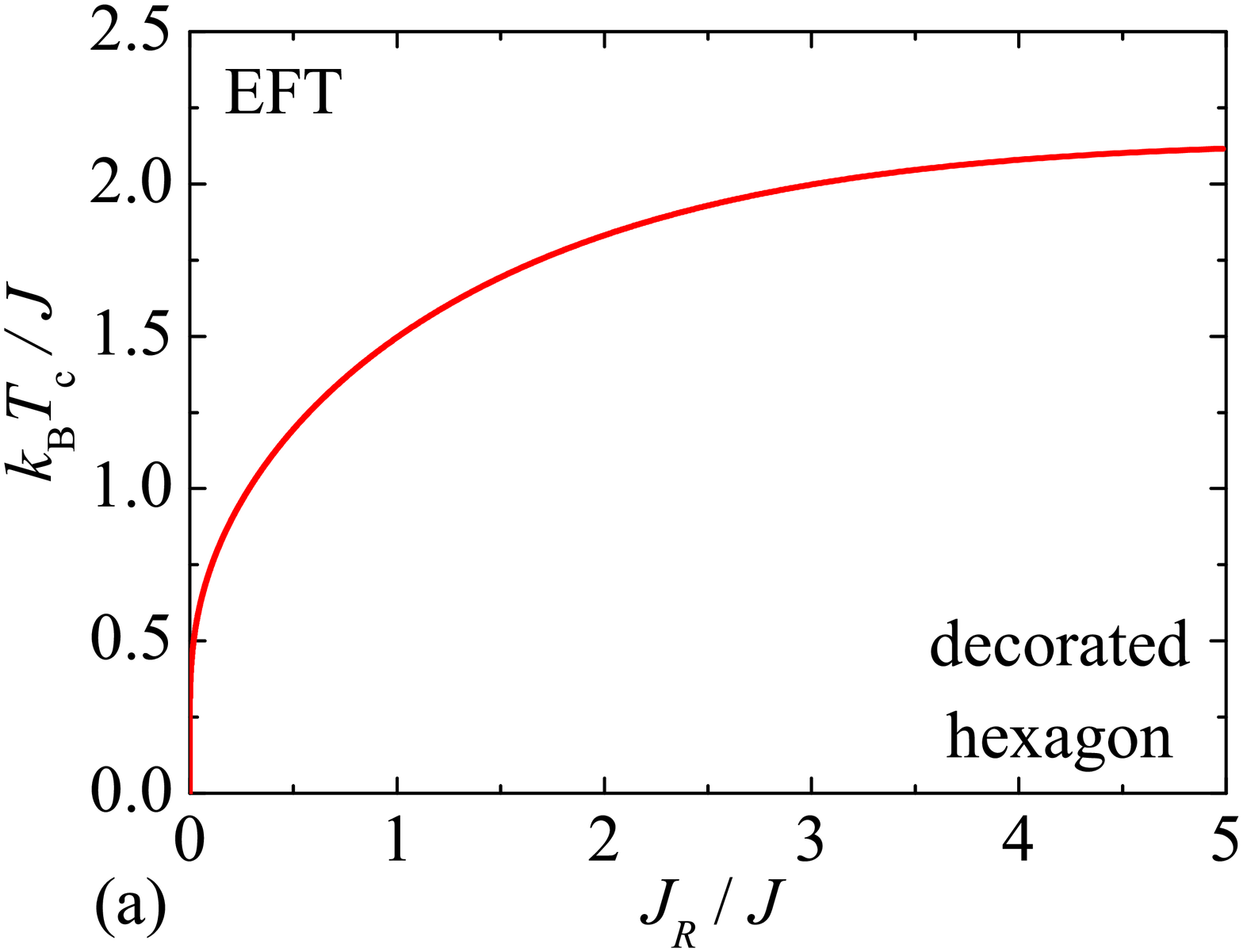}
\includegraphics[width=0.9\columnwidth]{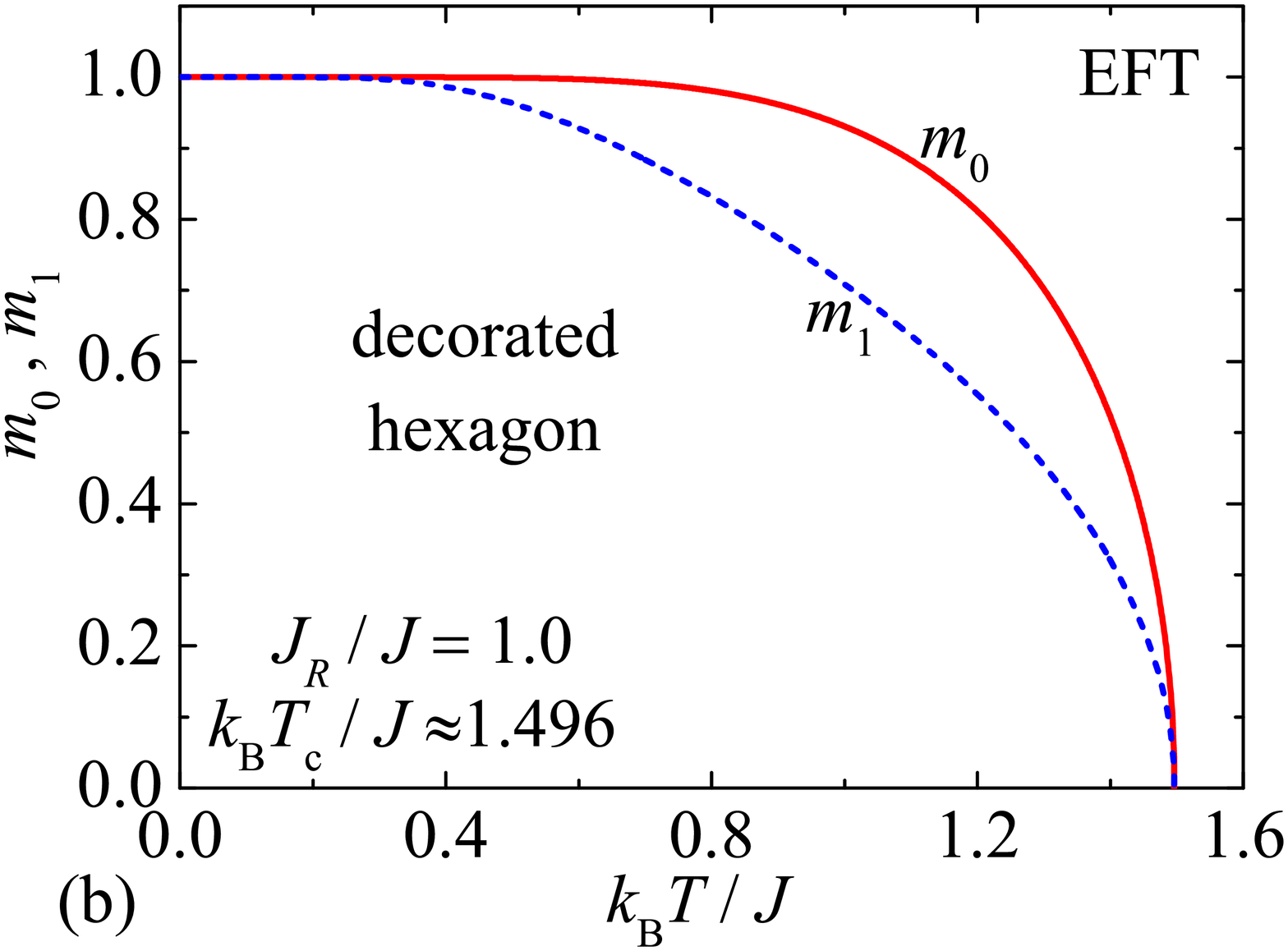}
\end{center}
\vspace*{-0.5cm}
\caption{(a) The critical temperature of the spin-1/2 Ising decorated hexagonal nanoparticle as calculated within the EFT; (b) Temperature variations of the spontaneous local magnetizations calculated for the spin-1/2 Ising decorated hexagonal nanoparticle within the EFT for the coupling ratio $J_R/J = 1$.}
\label{figdhtcm}
\end{figure}

The local magnetizations of the nodal and decorating spins of the spin-1/2 Ising decorated nanoparticle given by the Hamiltonian (\ref{eq:hamd}) can be calculated from the exact Callen-Suzuki spin identities \cite{call63,suzu65,balc02}:
\begin{eqnarray}
m_0 \!\!\!&\equiv&\!\!\! \langle S_j \rangle 
= \langle \tanh[\beta J (S_{j-1} + S_{j+1}) + \beta J_R \sigma_j] \rangle, \nonumber \\
m_1 \!\!\!&\equiv&\!\!\! \langle \sigma_j \rangle = \langle \tanh(\beta J_R S_j) \rangle, \qquad (j = 1-6). 
\label{eq:dmag}
\end{eqnarray}  
The couple of equations (\ref{eq:dmag}) can be rewritten using the differential operator technique to the  following form:
\begin{eqnarray}
m_0 \!\!\!&=&\!\!\! \langle \exp[(S_{j-1} + S_{j+1}) \nabla_x + \sigma_j \nabla_y] \rangle 
\tanh(\beta J x + \beta J_R y), \nonumber \\
m_1 \!\!\!&=&\!\!\! \langle \exp(S_j \nabla_x) \rangle \tanh(\beta J_R x). 
\label{eq:dmdo}
\end{eqnarray}
The subsequent application of the exact van der Waerden spin identity \cite{waer41}, the differential operator technique and the HK decoupling scheme \cite{kane93} for the higher-order correlations gives the following result:
\begin{eqnarray}
m_0 \!\!\!&=&\!\!\! 2 m_0 L_1 + m_1 L_2 + m_0^2 m_1 L_3, \nonumber \\
m_1 \!\!\!&=&\!\!\! m_0 L_0, 
\label{eq:dmvw}
\end{eqnarray} 
where the coefficients $L_0-L_3$ are defined as follows:
\begin{eqnarray}
L_0 \!\!\!\!\!&=&\!\!\!\!\! \tanh(\beta J_R), \nonumber \\
L_1 \!\!\!\!\!&=&\!\!\!\!\! \frac{1}{4} \! \left[\tanh (2\beta J \!+\! \beta J_R) 
             \!+\! \tanh (2\beta J \!-\! \beta J_R) \right]\!,  \nonumber \\
L_2 \!\!\!\!\!&=&\!\!\!\!\! \frac{1}{4} \! \left[\tanh (2\beta J \!+\! \beta J_R) 
             \!-\! \tanh (2\beta J \!-\! \beta J_R) 
             \!+\! 2 \tanh (\beta J_R) \right]\!,  \nonumber \\
L_3 \!\!\!\!\!&=&\!\!\!\!\! \frac{1}{4} \! \left[\tanh (2\beta J \!+\! \beta J_R) 
           \!-\! \tanh (2\beta J \!-\! \beta J_R) 
           \!-\! 2 \tanh (\beta J_R) \right]\!. \nonumber \\
\label{eq:dc}
\end{eqnarray} 
The elimination of either the local magnetization $m_0$ or $m_1$ from the set of two equations (\ref{eq:dmvw}) affords the following result for the spontaneous local magnetizations of the spin-1/2 Ising decorated hexa\-gonal nanoparticle:
\begin{eqnarray}
m_0 \!\!\!&=&\!\!\! \sqrt{\frac{1 - 2 L_1 - L_0 L_2}{L_0 L_3}}, \nonumber \\
m_1 \!\!\!&=&\!\!\! \sqrt{\frac{L_0 (1 - 2 L_1 - L_0 L_2)}{L_3}}.
\label{eq:dm}
\end{eqnarray}
It is obvious from Eq. (\ref{eq:dm}) that both spontaneous magnetizations $m_0$ and $m_1$ vanish at the critical temperature unambiguously determined by the critical condition $2 L_1 + L_0 L_2 = 1$. The critical temperature calculated from the aforementioned critical condition is plotted in Fig. \ref{figdhtcm}(a) as a function of the coupling ratio $J_R/J$, whereas this plot is in a perfect agreement with the results previously reported in Fig. 2 of Ref. \cite{kane21}. As one can see, the critical temperature monotonically decreases with a decline of the interaction ratio $J_R/J$ until it completely vanishes in the asymptotic limit $J_R/J \to 0$. This result would imply that the EFT predicts a spontaneous long-range order for the Ising nanoparticle just if it involves spins with the coordination number greater than two, because the investigated decorated nanoparticle decomposes in the limiting case $J_R/J = 0$ into the closed Ising chain involving spins with the coordination number two and  noninteracting spins. A presence of the spontaneous long-range order of the spin-1/2 Ising decorated hexagonal nanoparticle can be evidenced also by temperature dependences of the spontaneous magnetizations $m_0$ and $m_1$, which are depicted in Fig. \ref{figdhtcm}(b) for the specific case $J_R/J = 1$ serving for illustration. The local magnetization of the nodal spins $m_0$ is in general more robust against thermal fluctuations than the local magnetization of the decorating spins $m_1$ due to a higher coordination number of the nodal spins in comparison with the decorating ones.  

The internal energy of the spin-1/2 Ising decorated nanoparticle can be obtained from the formula 
$U = -NJ \langle S_j S_{j+1} \rangle - N J_R \langle S_j \sigma_j \rangle$, which includes two pair correlators easily calculable using the generalized Callen-Suzuki identities \cite{call63,suzu65,balc02}:
\begin{eqnarray}
\langle S_j S_{j+1} \rangle \!\!\!&=&\!\!\!
\langle S_{j+1} \tanh[\beta J (S_{j-1} + S_{j+1}) + \beta J_R \sigma_j] \rangle, \nonumber \\
\langle S_j \sigma_j \rangle \!\!\!&=&\!\!\! 
\langle \sigma_j \tanh[\beta J (S_{j-1} + S_{j+1}) + \beta J_R \sigma_j] \rangle. 
\label{eq:dcor}
\end{eqnarray}  
After taking advantage of the differential operator technique, the exact van der Waerden identity \cite{waer41} and the HK decoupling scheme \cite{kane93} introduced within the standard formulation of the EFT  one arrives at the following final result for the pair correlators expressed in terms of the local magnetizations (\ref{eq:dm}):
\begin{eqnarray}
\langle S_j S_{j+1} \rangle \!\!\!&=&\!\!\! (1 + m_0^2) L_1 + m_0 m_1 (L_2 + L_3), \nonumber \\
\langle S_j \sigma_j \rangle \!\!\!&=&\!\!\! 2 m_0 m_1 L_1 + L_2 + m_0^2 L_3. 
\label{eq:dcorf}
\end{eqnarray}

\subsubsection{Exact results} 

The spin-1/2 Ising decorated nanoparticle given by the Hamiltonian (\ref{eq:hamd}) can be alternatively viewed as the spin-1/2 Ising branched chain, which has on each site lateral branching involving one decorating spin and is being considered under the periodic boundary condition [see Fig. \ref{fig1d}(a)]. Hence, it follows that the exact solution for the spin-1/2 Ising decorated nanoparticle (\ref{eq:hamd}) can be readily derived with the help of transfer-matrix approach \cite{kram41}. The partition function of the spin-1/2 Ising decorated nanoparticle can be substantially simplified when performing at first a summation over spin states of the decorating spins:
\begin{eqnarray}
Z \!\!\!&=&\!\!\! \sum_{ \{S\} } \sum_{ \{\sigma\} } \exp(-\beta H) \nonumber \\
  \!\!\!&=&\!\!\! \sum_{ \{S\} } \prod_{j=1}^{N} \left[\exp(\beta J S_j S_{j+1}) 
	\sum_{\sigma_j = \pm 1} \exp(\beta J_R \sigma_j S_j) \right] \nonumber \\
  \!\!\!&=&\!\!\! 2^N \cosh^N(\beta J_R) \sum_{ \{S\} } \prod_{j=1}^{N} \exp(\beta J S_j S_{j+1}) \nonumber \\
	\!\!\!&=&\!\!\! 2^N \cosh^N(\beta J_R) \sum_{ \{S\} } \prod_{j=1}^{N} {\rm T} \, (S_j, S_{j+1}).
\label{pfd}
\end{eqnarray}
Note that the indicated summations $\sum_{ \{S\} }$ and $\sum_{ \{\sigma\} }$ are carried out over all available spin states of the nodal and decorating spins, respectively. In the last line of Eq. (\ref{pfd}) we have identified the expression ${\rm T} \, (S_j, S_{j+1}) = \exp(\beta J S_j S_{j+1})$ with two-by-two transfer matrix:
\begin{eqnarray}
{\rm T} \, (S_j, S_{j+1}) = \left( \begin{array}{cc}
                           \exp(\beta J) & \exp(-\beta J) \\ 
													 \exp(-\beta J) & \exp(\beta J)  
                          \end{array} 
                  \right).	
\label{tmd}
\end{eqnarray}
The consecutive summation over states of the nodal spins in Eq. (\ref{pfd}) allows one to express the partition function in terms of a trace of $N$th power of the transfer matrix (\ref{tmd}):
\begin{eqnarray}
Z = 2^N \cosh^N (\beta J_R) \mbox{Tr}\, T^N = 2^N \cosh^N (\beta J_R) (\lambda_1^N + \lambda_2^N),
\label{pfdtm}
\end{eqnarray}
which is in turn given due to a trace invariance through two eigenvalues $\lambda_1$ and $\lambda_2$ of the transfer matrix (\ref{tmd}). A straightforward diagonalization of the transfer matrix (\ref{tmd}) yields the eigenvalues $\lambda_1 = 2 \cosh(\beta J)$ and $\lambda_2 = 2 \sinh(\beta J)$ what in fact completes the exact calculation of the partition function:
\begin{eqnarray}
Z = 4^N \cosh^N (\beta J_R) [\cosh^N (\beta J) + \sinh^N (\beta J)].
\label{pfdf}
\end{eqnarray}
It is worthwhile to remark that the partition function of the spin-1/2 Ising decorated nanoparticle is determined by both transfer-matrix eigenvalues $\lambda_1 = 2 \cosh(\beta J)$ and $\lambda_2 = 2 \sinh(\beta J)$ in contrast to the infinite branched Ising chain shown in Fig. \ref{fig1d}(a), whose magnetic behavior is governed in the thermodynamic limit $N \to \infty$ just by the largest transfer-matrix eigenvalue $\lambda_1 = 2 \cosh(\beta J)$. 

\begin{figure}[t]
\begin{center}
\includegraphics[width=0.9\columnwidth]{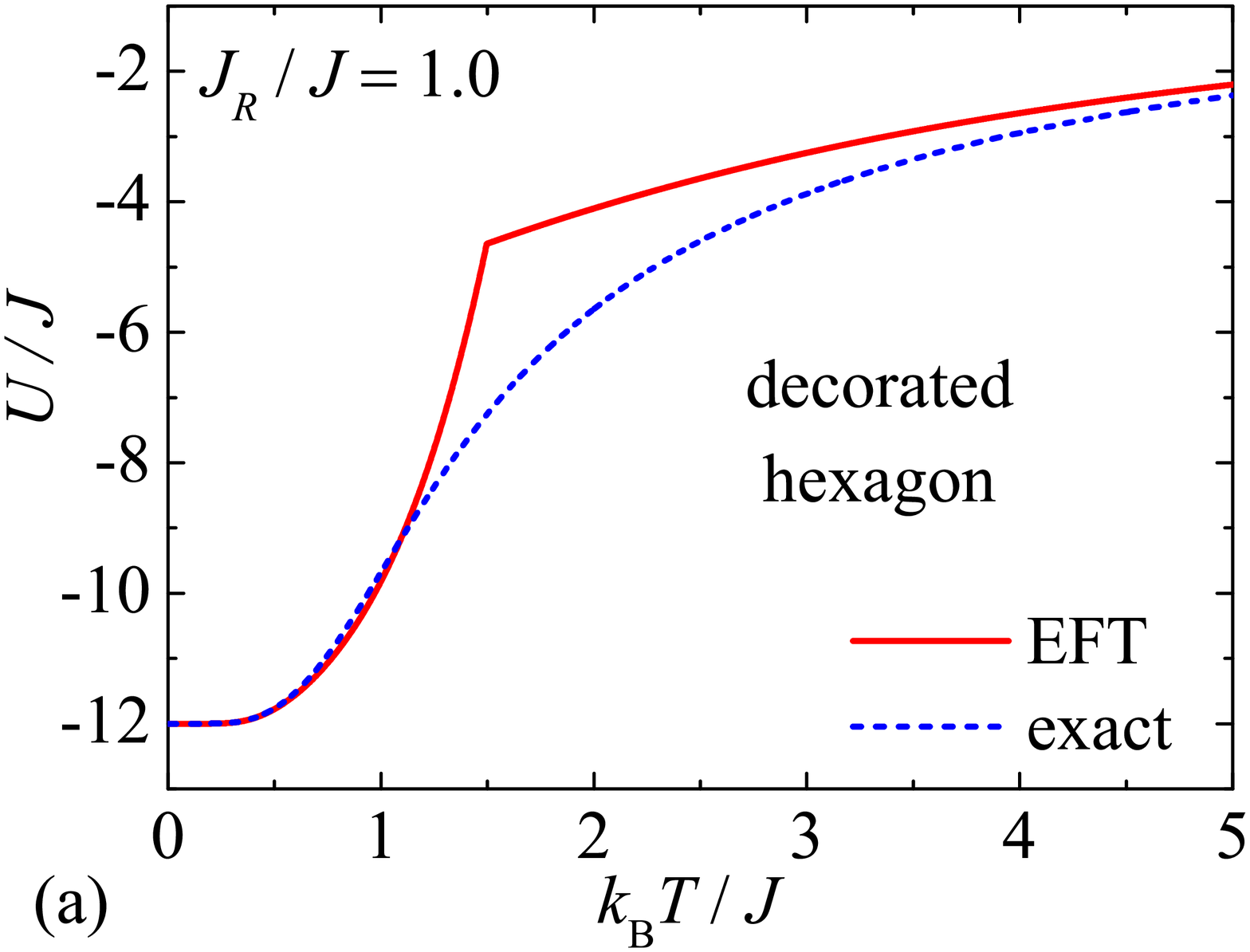}
\includegraphics[width=0.9\columnwidth]{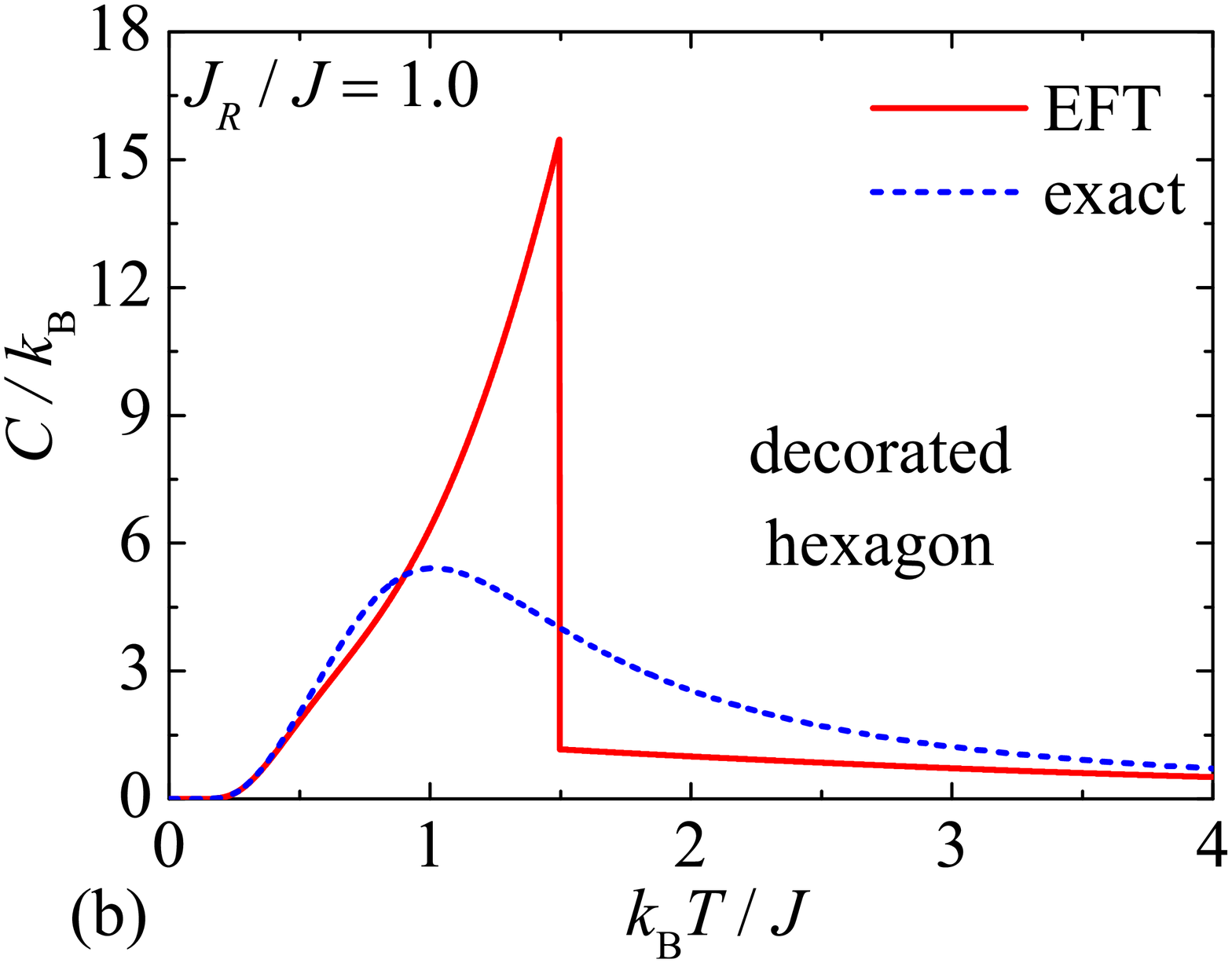}
\end{center}
\vspace*{-0.5cm}
\caption{Temperature variations of the internal energy (panel a) and specific heat (panel b) calculated for the spin-1/2 Ising decorated hexagonal nanoparticle for the specific value of the interaction ratio $J_R/J = 1$ within the EFT (solid lines) and exact method (broken lines).}
\label{figdhuc}
\end{figure}

Importantly, it can be easily proved that the spontaneous local magnetizations of the spin-1/2 Ising decorated nanoparticle identically equal zero for the nodal as well as decorating spins:
\begin{eqnarray}
m_0 \!\!\!&\equiv&\!\!\! \langle S_j \rangle = \frac{1}{Z} \sum_{ \{S\} } \sum_{ \{\sigma\} } S_j \exp(-\beta H) = 0,  \nonumber \\
m_1 \!\!\!&\equiv&\!\!\! \langle \sigma_j \rangle = \frac{1}{Z} \sum_{ \{S\} } \sum_{ \{\sigma\} } \sigma_j \exp(-\beta H) = 0.
\label{dme}
\end{eqnarray}
Owing to this fact, the nonzero spontaneous magnetizations depicted in Fig. \ref{figdhtcm}(b) according to Eq. (\ref{eq:dm}) are repeatedly just artifacts of the HK decoupling approximation exploited within the EFT \cite{kane93}. To verify absence of a spontaneous long-range order of the spin-1/2 Ising decorated nanoparticle one may employ the exact result (\ref{pfdf}) for the partition function in order to calculate the internal energy $U = -\partial \ln Z/\partial \beta$ and the specific heat $C = \partial U/\partial T$. In such a way obtained  exact results for the internal energy and specific heat of the spin-1/2 Ising decorated hexagonal nanoparticle with $N=6$ are displayed in Fig. \ref{figdhuc} together with the corresponding results of the EFT. The results derived within the EFT and transfer-matrix method are in a good accordance in the low- and high-temperature region, while they contradict themselves at moderate temperatures where the EFT predicts a continuous phase transition at the critical temperature $k_B T_c/J \approx 1.496$ for the specific case $J_R/J = 1$. The EFT repeatedly implies presence of a cusp in the temperature dependence of the internal energy and a finite jump in the temperature dependence of the specific heat, which are in obvious contrast with smooth continuous temperature variations of the internal energy and specific heat excluding existence of a finite-temperature phase transition and spontaneous long-range order. The exact results again disqualify feasibility of the EFT for the analysis of magnetic and thermodynamic properties of this zero-dimensional Ising spin cluster.  

\subsection{Star of David}
\label{sec:david}

Another paradigmatic example of our particular interest is the spin-1/2 Ising bond-decorated nanoparticle given by the Hamiltonian:
\begin{eqnarray}
H = - J \sum_{j=1}^N S_j S_{j+1} - J_R \sum_{j=1}^N \sigma_{j} (S_j + S_{j+1}),
\label{eq:hamsod}
\end{eqnarray}
which involves the Ising spins $S_j = \pm 1$ and $\sigma_j = \pm 1$ assigned to nodal and decorating sites of the nanoparticle under the periodic boundary condition $S_{N+1} \equiv S_1$. Although the solution for the spin-1/2 Ising bond-decorated nanoparticle given by the Hamiltonian (\ref{eq:hamsod}) formulated within the EFT and transfer-matrix method can be obtained for the quite general number of spins $N \geq 3$, we will be particularly interested in the special case with $N=6$ corresponding to the spin-1/2 Ising cluster with the geometric shape of the star of David shown in Fig. \ref{fig0d}(d). Note furthermore that the spin-1/2 Ising bond-decorated nanoparticle given by the Hamiltonian (\ref{eq:hamsod}) corresponds in the thermodynamic limit $N \to \infty$ to a spin-1/2 Ising sawtooth ($\Delta$) chain schematically shown in Fig. \ref{fig1d}(b).

\subsubsection{Effective-field theory} 

\begin{figure}[t]
\begin{center}
\includegraphics[width=0.9\columnwidth]{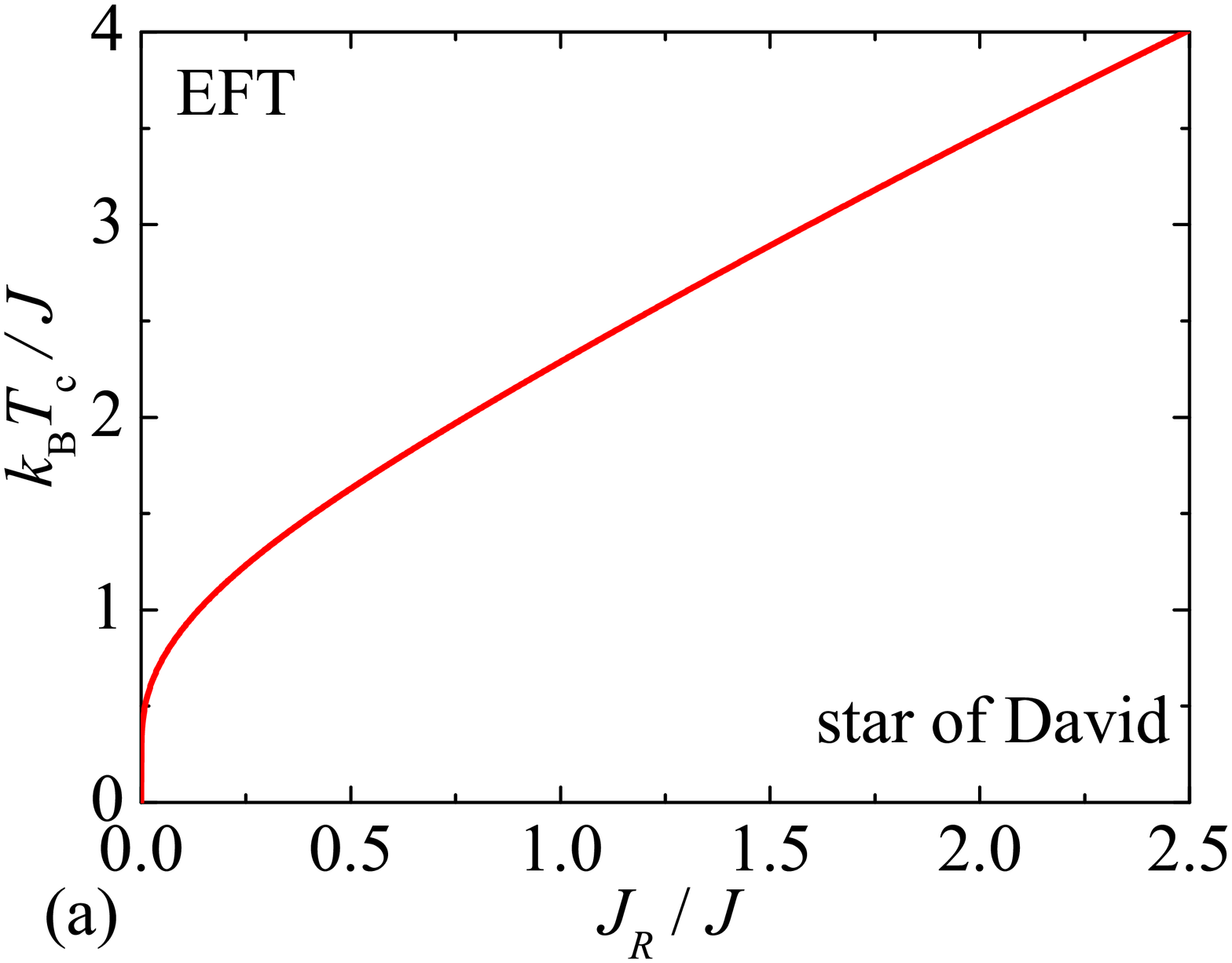}
\includegraphics[width=0.9\columnwidth]{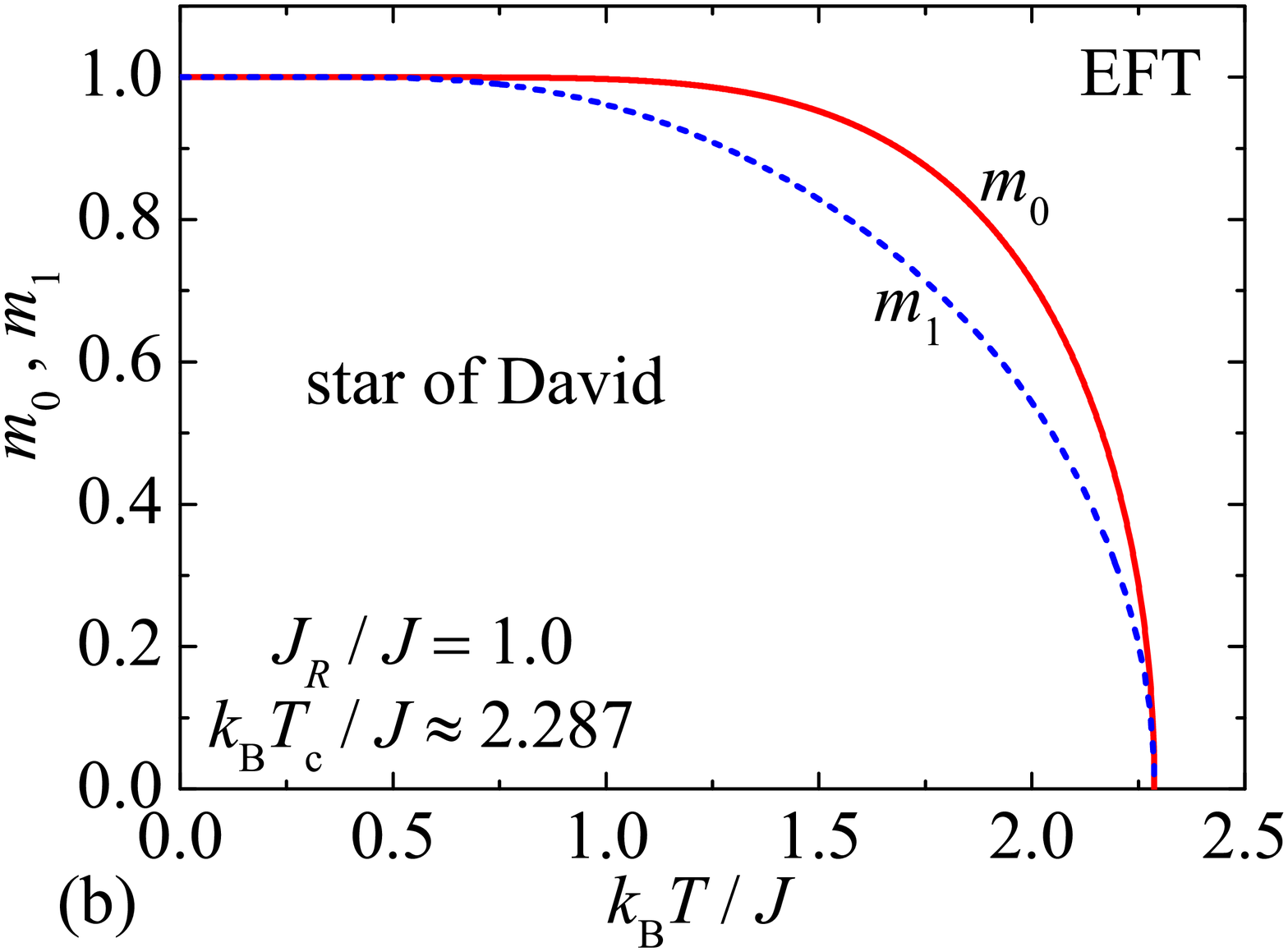}
\end{center}
\vspace*{-0.5cm}
\caption{(a) The critical temperature of the spin-1/2 Ising star of David as calculated within the EFT; (b) Temperature variations of the spontaneous local magnetizations calculated for the spin-1/2 Ising star of David within the EFT for the particular value of the coupling ratio $J_R/J = 1$.}
\label{figsodtcm}
\end{figure}

The local magnetizations of the nodal and decorating spins of the spin-1/2 Ising star of David can be obtained within the EFT from the exact Callen-Suzuki spin identities \cite{call63,suzu65,balc02}:
\begin{eqnarray}
m_0 \!\!\!&\equiv&\!\!\! \langle S_j \rangle 
= \langle \tanh[\beta J (S_{j-1} + S_{j+1}) + \beta J_R (\sigma_{j-1} + \sigma_j)] \rangle, \nonumber \\
m_1 \!\!\!&\equiv&\!\!\! \langle \sigma_j \rangle = \langle \tanh[\beta J_R (S_j + S_{j+1})] \rangle, 
\quad (j = 1-6). 
\label{eq:sodmag}
\end{eqnarray}  
The exact spin identities (\ref{eq:sodmag}) can be rewritten with the help of the differential operator technique as follows:
\begin{eqnarray}
m_0 \!\!\!\!\!&=&\!\!\!\!\! \langle \exp[(S\!_{j-1} \!\!+\! S\!_{j+1}) \nabla_x \!+\! 
(\sigma\!_{j-1} \!\!+\! \sigma\!_j) \nabla_y] \rangle \! \tanh(\beta J x \!+\! \beta J_R y), \nonumber \\
m_1 \!\!\!\!\!&=&\!\!\!\!\! \langle \exp[(S_j + S_{j+1}) \nabla_x] \rangle \tanh(\beta J_R x). 
\label{eq:sodmdo}
\end{eqnarray}
Using the exact van der Waerden spin identity \cite{waer41}, the differential operator and the HK decoupling scheme \cite{kane93} introduced within the standard formulation of the EFT provides the following couple of mutually inter-connected equations for the local magnetizations:
\begin{eqnarray}
m_0 \!\!\!&=&\!\!\! 2 m_0 N_1 + 2 m_1 N_2 + 2 m_0^2 m_1 N_3 + 2 m_0 m_1^2 N_4, \nonumber \\
m_1 \!\!\!&=&\!\!\! m_0 N_0, 
\label{eq:sodmvw}
\end{eqnarray} 
where the coefficients $N_0-N_4$ are defined as follows:
\begin{eqnarray}
N_0 \!\!\!\!\!\!&=&\!\!\!\!\!\! \tanh(2 \beta J_R), \nonumber \\
N_1 \!\!\!\!\!\!&=&\!\!\!\!\!\! \frac{1}{8} \! \left[\tanh (2\beta J \!\!+\!\! 2\beta J_R) 
             \!+\! \tanh (2\beta J \!\!-\!\! 2 \beta J_R) \!+\! 2 \! \tanh (2 \beta J) \right]\!,  \nonumber \\
N_2 \!\!\!\!\!\!&=&\!\!\!\!\!\! \frac{1}{8} \! \left[\tanh (2\beta J \!\!+\!\! 2\beta J_R) 
             \!-\! \tanh (2\beta J \!\!-\!\! 2 \beta J_R) \!+\! 2 \! \tanh (2 \beta J_R) \right]\!,  \nonumber \\
N_3 \!\!\!\!\!\!&=&\!\!\!\!\!\! \frac{1}{8} \! \left[\tanh (2\beta J \!\!+\!\! 2\beta J_R) 
             \!-\! \tanh (2\beta J \!\!-\!\! 2 \beta J_R) \!-\! 2 \! \tanh (2 \beta J_R) \right]\!,  \nonumber \\
N_4 \!\!\!\!\!\!&=&\!\!\!\!\!\! \frac{1}{8} \! \left[\tanh (2\beta J \!\!+\!\! 2\beta J_R) 
             \!+\! \tanh (2\beta J \!\!-\!\! 2 \beta J_R) \!-\! 2 \! \tanh (2 \beta J) \right]\!,  \nonumber \\
\label{eq:sodc}
\end{eqnarray} 
Eliminating either the local magnetization $m_0$ or $m_1$ from Eqs. (\ref{eq:sodmvw}) provides the following expression for the spontaneous magnetizations of the spin-1/2 Ising star of David:
\begin{eqnarray}
m_0 \!\!\!&=&\!\!\! \sqrt{\frac{1 - 2 N_1 - 2 N_0 N_2}{2 N_0 N_3 + 2 N_0^2 N_4}}, \nonumber \\
m_1 \!\!\!&=&\!\!\! \sqrt{\frac{N_0^2 (1 - 2 N_1 - 2 N_0 N_2)}{2 N_0 N_3 + 2 N_0^2 N_4}}.
\label{eq:sodm}
\end{eqnarray}
It can be easily understood from Eq. (\ref{eq:sodm}) that both spontaneous magnetizations $m_0$ and $m_1$ tend to zero whenever the critical condition $2 N_1 + 2 N_0 N_2 = 1$ is met. The critical temperature derived from this critical condition is depicted in Fig. \ref{figsodtcm}(a) against the interaction ratio $J_R/J$. If the relative strength of the coupling constants is sufficiently large $J_R/J \gtrsim 1$, then, the critical temperature displays a linear rise with increasing of the interaction ratio $J_R/J$, while it shows a steep decline down to zero as the interaction ratio vanishes $J_R/J \to 0$. The spin-1/2 Ising star of David decomposes in the limiting case $J_R/J = 0$ into the closed Ising chain involving spins with the coordination number two and noninteracting spins and hence, the EFT only predicts a spontaneous long-range order just for $J_R/J \neq 0$ when the spin-1/2 Ising star of David includes spins with the coordination number four. The spontaneous long-range order of the spin-1/2 Ising star of David is consistent with temperature dependences of the spontaneous magnetizations $m_0$ and $m_1$, which are plotted in Fig. \ref{figsodtcm}(b) for the specific case $J_R/J = 1$ serving for illustration. As one can see, the local magnetization of the nodal spins $m_0$ with the coordination number four is more resistant with respect to thermal fluctuations than the local magnetization of the decorating spins $m_1$ with the coordination number two.  

The internal energy of the spin-1/2 Ising star of David depends on two pair correlators $U = -NJ \langle S_j S_{j+1} \rangle - 2 N J_R \langle S_j \sigma_j \rangle$, which can be calculated from the generalized Callen-Suzuki identities \cite{call63,suzu65,balc02}:
\begin{eqnarray}
\langle S_j S_{j+1} \rangle \!\!\!\!\!&=&\!\!\!\!\!
\langle S_{j+1} \tanh[\beta J (S_{j-1} \!+\! S_{j+1}) \!+\! \beta J_R (\sigma_{j-1} \!+\! \sigma_j)] \rangle, \nonumber \\
\langle S_j \sigma_j \rangle \!\!\!\!\!&=&\!\!\!\!\!
\langle \sigma_j \tanh[\beta J (S_{j-1} \!+\! S_{j+1}) \!+\! \beta J_R (\sigma_{j-1} \!+\! \sigma_j)] \rangle, \nonumber \\
\langle S_j \sigma_j \rangle \!\!\!\!\!&=&\!\!\!\!\! 
\langle S_j \tanh[\beta J_R (S_{j} \!+\! S_{j+1})] \rangle.
\label{eq:sodcor}
\end{eqnarray}  
By making use of the differential operator, the exact van der Waerden identity \cite{waer41} and the HK decoupling scheme \cite{kane93} within the EFT one may express the pair correlators in terms of the local magnetizations (\ref{eq:sodm}):
\begin{eqnarray}
\langle S_j \sigma_j \rangle \!\!\!\!\!\!&=&\!\!\!\!\!\! \frac{(1 \!\!+\!\! m_0^2) N_{5} \!\!+\!\! m_0 m_1 N_8 \!\!+\!\! (m_1^2 \!\!-\!\! m_0^2) N_{7} \!\!-\!\! (1 \!\!+\!\! m_1^2) N_{6} \!\!+\!\! N_3}{N_4(N_4 \!\!+\!\! N_1 \!\!-\!\! 1) \!\!-\!\! N_3 (N_3 \!+\! N_2 \!-\! 2/N_0)},  \nonumber \\
\langle S\!_j S\!_{j+1} \rangle \!\!\!\!\!\!&=&\!\!\!\!\!\! \frac{2}{N_0} \langle S_j \sigma_j \rangle \!-\! 1. 
\label{eq:sodcorf}
\end{eqnarray}
Above, we have introduced the following notation for the newly defined coefficients: $N_{5} = N_1 N_3$, $N_{6} = N_2 N_4$, $N_{7} = N_3 N_4$, $N_8 = N_3 (N_2 + N_3) - N_4(N_1 + N_4)$.

\subsubsection{Exact results} 

The partition function of the spin-1/2 Ising bond-decorated nanoparticle given by the Hamiltonian (\ref{eq:hamsod}), and its particular case with $N=6$ corresponding to the star of David, can be rigorously calculated for instance by the exact enumeration of states or the transfer-matrix method \cite{kram41}. The antiferromagnetic version of the spin-1/2 Ising star of David has been examined in detail by the exact enumeration of states in Ref. \cite{zuko18}. In what follows, we will solve the spin-1/2 Ising bond-decorated nanoparticle within the transfer-matrix method \cite{kram41}, which allows the exact solution for arbitrary system size including $N=6$ pertinent to the star of David or even $N \to \infty$ corresponding to the spin-1/2 sawtooth ($\Delta$) chain [see Fig. \ref{fig1d}(b)]. Before the rigorous concept of the transfer-matrix approach will be applied it is convenient at first to perform a summation over spin states of the decorating spins:
\begin{eqnarray}
Z \!\!\!&=&\!\!\! \sum_{ \{S\} } \sum_{ \{\sigma\} } \exp(-\beta H) \nonumber \\
  \!\!\!&=&\!\!\! \sum_{ \{S\} } \prod_{j=1}^{N} \left\{\exp(\beta J S_j S_{j+1}) \!\!\!
	\sum_{\sigma_j = \pm 1} \!\!\! \exp[\beta J_R \sigma_j (S_j + S_{j+1})] \right\} \nonumber \\
  \!\!\!&=&\!\!\! \sum_{ \{S\} } \prod_{j=1}^{N} \left\{2 \exp(\beta J S_j S_{j+1}) 
	                \cosh [\beta J_R (S_j + S_{j+1})] \right\} \nonumber \\
	\!\!\!&=&\!\!\! \sum_{ \{S\} } \prod_{j=1}^{N} {\rm T} \, (S_j, S_{j+1}).
\label{pfsod}
\end{eqnarray}
The Boltzmann weight ${\rm T} \, (S_j, S_{j+1}) = 2 {\rm e}^{\beta J S_j S_{j+1}} \cosh [\beta J_R (S_j + S_{j+1})]$ introduced in the last line of Eq. (\ref{pfsod}) can be identified with two-by-two transfer matrix:
\begin{eqnarray}
{\rm T} \, (S_j, S_{j+1}) = \left( \!\begin{array}{cc}
                          \! 2 {\rm e}^{\beta J} \! \cosh(2 \beta J_R) & 2 {\rm e}^{-\beta J} \\ 
													 2 {\rm e}^{-\beta J} & \! 2 {\rm e}^{\beta J} \! \cosh(2 \beta J_R) 
                          \end{array} 
                  \! \right)\!.	
\label{tmsod}
\end{eqnarray}
The consecutive summation over states of the nodal spins in Eq. (\ref{pfsod}) is equivalent to multiplication of the relevant transfer matrices what allows one to express the partition function in terms of a trace of $N$th power of the transfer matrix (\ref{tmsod}):
\begin{eqnarray}
Z = \sum_{ \{S\} } \prod_{j=1}^{N} {\rm T} \, (S_j, S_{j+1}) = \mbox{Tr}\, {\rm T}^N = \lambda_1^N + \lambda_2^N,
\label{pfsodtm}
\end{eqnarray}
which is given by two eigenvalues of the transfer matrix $\lambda_1 = 2 {\rm e}^{\beta J} \cosh(2 \beta J_R) + 2 {\rm e}^{-\beta J}$ and $\lambda_2 = 2 {\rm e}^{\beta J}\cosh(2 \beta J_R) - 2 {\rm e}^{-\beta J}$. Substituting the transfer-matrix eigenvalues into the partition function (\ref{pfsodtm}) actually completes its exact calculation:
\begin{eqnarray}
Z \!\!\!&=&\!\!\! [2 \exp(\beta J) \cosh(2 \beta J_R) + 2 \exp(-\beta J)]^N \nonumber \\
  \!\!\!&+&\!\!\! [2 \exp(\beta J) \cosh(2 \beta J_R) - 2 \exp(-\beta J)]^N.
\label{pfsodf}
\end{eqnarray}
The partition function of the spin-1/2 Ising star of David can be obtained from Eq. (\ref{pfsodf}) by inserting therein the specific value $N=6$. It is noteworthy that both transfer-matrix eigenvalues $\lambda_1$ and $\lambda_2$ determine the partition function for any finite number of $N$. Contrary to this, the exact result for the partition function of the spin-1/2 Ising sawtooth ($\Delta$) chain retrieved by the Hamiltonian (\ref{eq:hamsod}) in the thermodynamic limit $N \to \infty$ is governed just the larger transfer-matrix eigenvalue $\lambda_1$. 

\begin{figure}[t]
\begin{center}
\includegraphics[width=0.9\columnwidth]{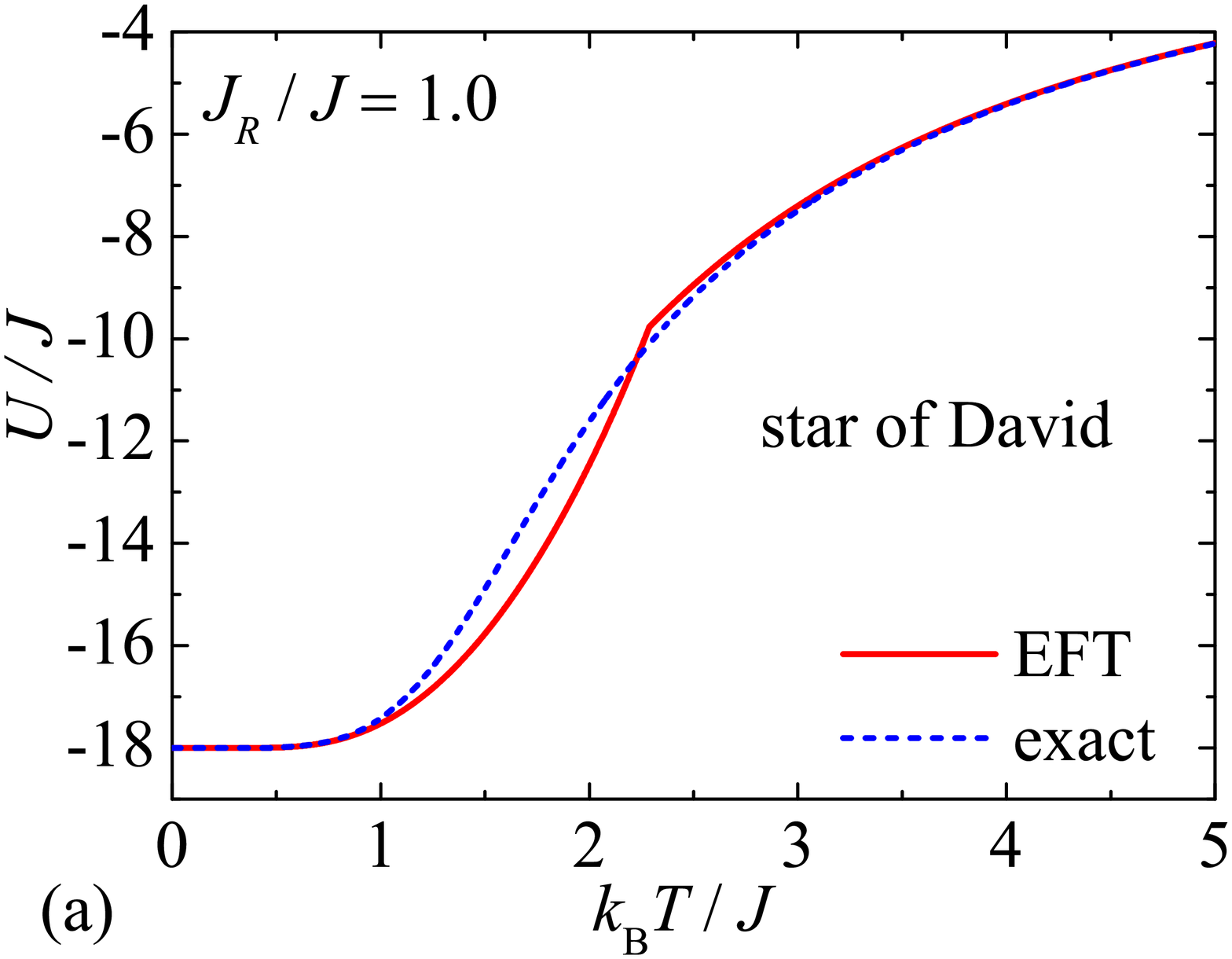}
\includegraphics[width=0.9\columnwidth]{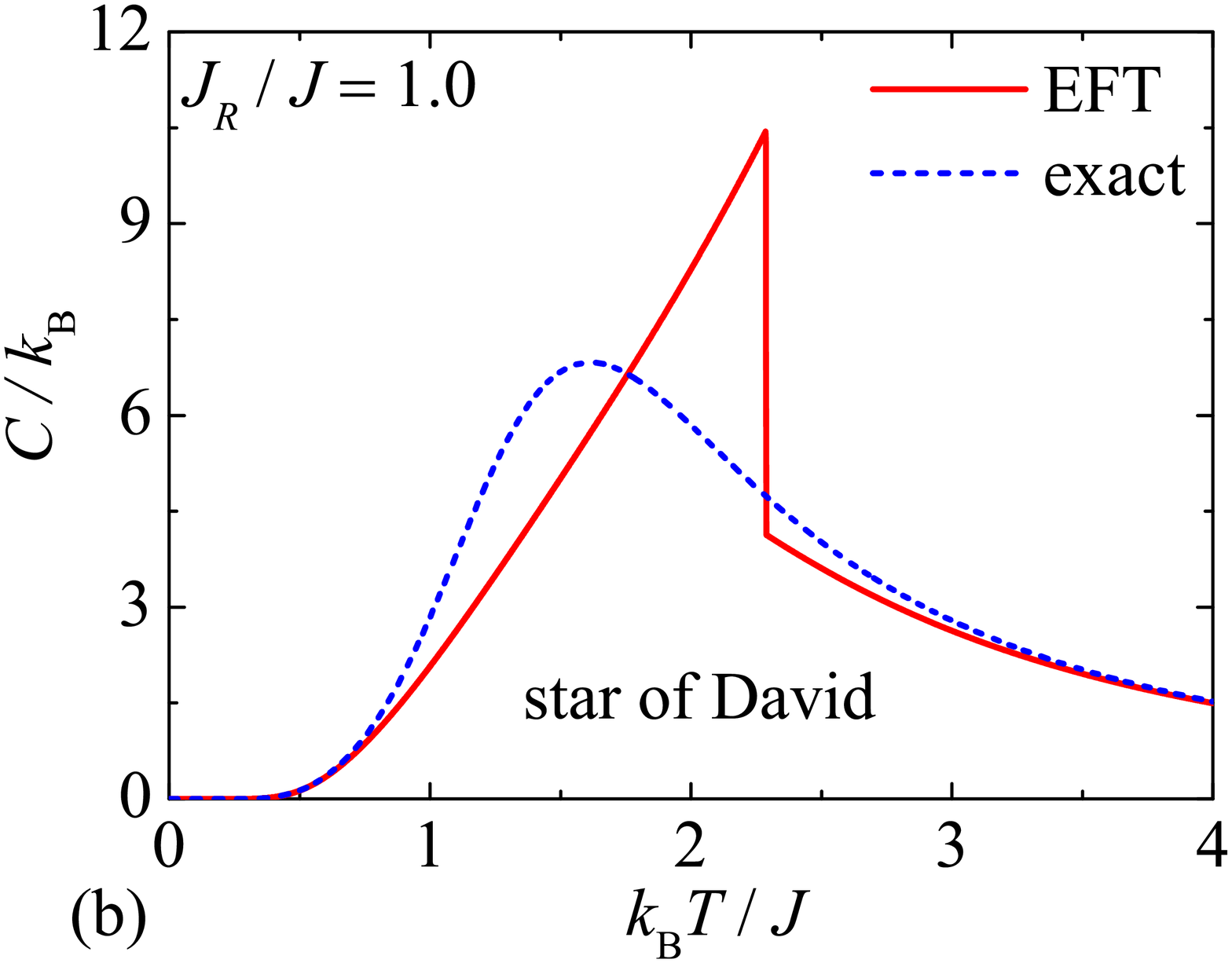}
\end{center}
\vspace*{-0.5cm}
\caption{Temperature variations of the internal energy (panel a) and specific heat (panel b) calculated for the spin-1/2 Ising star of David for the specific value of the interaction ratio $J_R/J = 1$ within the EFT (solid lines) and exact method (broken lines).}
\label{figsoduc}
\end{figure}

It can be easily verified that the spontaneous local magnetizations of the spin-1/2 Ising star of David equal zero for the nodal as well as decorating spins:
\begin{eqnarray}
m_0 \!\!\!&\equiv&\!\!\! \langle S_j \rangle = \frac{1}{Z} \sum_{ \{S\} } \sum_{ \{\sigma\} } S_j \exp(-\beta H) = 0,  \nonumber \\
m_1 \!\!\!&\equiv&\!\!\! \langle \sigma_j \rangle = \frac{1}{Z} \sum_{ \{S\} } \sum_{ \{\sigma\} } \sigma_j \exp(-\beta H) = 0,
\label{dmesod}
\end{eqnarray}
what is in sharp contrast with nonzero results (\ref{eq:sodm}) of the spontaneous magnetizations depicted in Fig. \ref{figsodtcm}(b) according to the EFT \cite{kane93}. The absence of a spontaneous long-range order of the spin-1/2 Ising star of David can be independently corroborated with the help of the exact result (\ref{pfsodf}) for the partition function, which allows a straightforward calculation of the internal energy $U = -\partial \ln Z/\partial \beta$ and the specific heat $C = \partial U/\partial T$. The exact results for the internal energy and specific heat of the spin-1/2 Ising star of David are displayed in Fig. \ref{figsoduc} together with the corresponding results of the EFT. Both approaches are in good agreement in the low- and high-temperature regions, while they contradict themselves at moderate temperatures where the EFT declares presence of a continuous phase transition in contrast with the exact results. According to the EFT, temperature dependences of the internal energy and specific heat display a cusp and a finite jump at the critical temperature $k_B T_c/J \approx 2.287$ for the specific case $J_R/J = 1$, while both quantities exhibit smooth continuous temperature changes precluding existence of a finite-temperature phase transition and spontaneous long-range order. The exact results thus repeatedly disqualify feasibility of the EFT for the analysis of magnetic and thermodynamic properties of this zero-dimensional Ising spin cluster.  

\section{One-dimensional Ising nanosystems}
\label{sec:3}

\begin{figure}[t]
\begin{center}
\includegraphics[width=0.8\columnwidth]{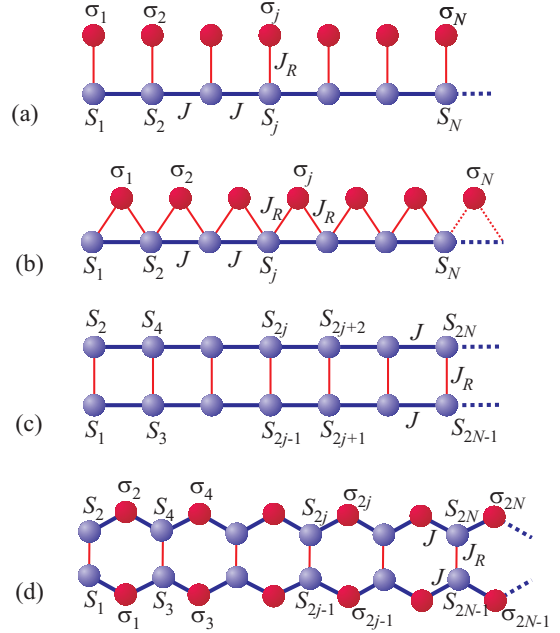}
\end{center}
\vspace*{-0.5cm}
\caption{A few typical examples of one-dimensional Ising spin nanosystems: (a) branched chain; (b) sawtooth ($\Delta$) chain; (c) two-leg ladder; (d) hexagonal ladder. Dotted lines illustrate periodic boundary conditions.}
\label{fig1d}
\end{figure}

It is quite evident from previous exact calculations that the partition functions of the spin-1/2 Ising branched chain [Fig. \ref{fig1d}(a)] and the spin-1/2 Ising sawtooth chain [Fig. \ref{fig1d}(b)] acquired from Eqs. (\ref{pfdf}) and (\ref{pfsodf}) in the thermodynamic limit $N \to \infty$ are smooth analytic functions of temperature and the same statement holds true also for any temperature derivative of them. These findings are thus in a perfect agreement with absence of finite-temperature phase transition and a spontaneous long-range order in those two one-dimensional Ising spin systems. In this section we will investigate other three paradigmatic examples of the one-dimensional Ising nanosystems, each of which will be further treated within the framework of the EFT and exact transfer-matrix method \cite{kram41}. In particular, we will consider the spin-1/2 Ising chain, two-leg and hexagonal ladders schematically illustrated in Fig. \ref{fig1d}(c) and (d), respectively. 

\subsection{Chain}
\label{sec:3a}

The spin-1/2 Ising chain is the famous example of one-dimensional lattice-statistical model given by the Hamiltonian:
\begin{equation}
H = -J \sum_{j=1}^N S_j S_{j+1},
\label{eq:hic}
\end{equation}
which can be treated by the EFT as well as several exact methods \cite{stre15}. For brevity, we will therefore review here only a few basic steps of both theoretical approaches.  

\subsubsection{Effective-field theory} 

The local magnetization of the spin-1/2 Ising chain can be calculated from the exact Callen-Suzuki spin identity \cite{call63,suzu65,balc02}:
\begin{eqnarray}
m \equiv \langle S_j \rangle = \langle \tanh[\beta J (S_{j-1} + S_{j+1})] \rangle, 
\label{eq:chmag}
\end{eqnarray}  
which can be rewritten using the differential operator technique to the following form:
\begin{eqnarray}
m = \langle \exp[(S_{j-1} + S_{j+1}) \nabla_x] \rangle \tanh(\beta J x). 
\label{eq:cmdo}
\end{eqnarray}
The subsequent application of the exact van der Waerden spin identity \cite{waer41} and the differential operator technique immediately provides the following result: 
\begin{eqnarray}
m = m \tanh (2 \beta J), 
\label{eq:chmvw}
\end{eqnarray} 
or equivalently:
\begin{eqnarray}
m [1 - \tanh (2 \beta J)] = 0.
\label{eq:ctc}
\end{eqnarray} 
It directly follows from Eq. (\ref{eq:ctc}) that the expression given in a square bracket should equal zero in order to have nonzero spontaneous magnetization $m \neq 0$. However, the condition $\tanh (2 \beta J) = 1$ restricts the nonzero spontaneous magnetization just to absolute zero temperature, which is simultaneously the critical temperature of the spin-1/2 Ising chain. A great success of the EFT with respect to the standard mean-field theory consists in predicting absence of a spontaneous long-range order of the spin-1/2 Ising chain is accordance with the exact solution presented hereafter \cite{kane93,stre15}. The reason for exactness of the approach based on the Callen-Suzuki spin identity \cite{call63,suzu65,balc02}, the differential operator and van der Waerden spin identity \cite{waer41} lies in an exactness of each given step, whereas the HK decoupling scheme \cite{kane93} as the only approximate step splitting higher-order correlations within the EFT  does not need to be applied in this particular case.

The internal energy of the spin-1/2 Ising chain can be expressed in terms of the nearest-neighbor pair correlator $U = -NJ \langle S_j S_{j+1} \rangle$, which can be also calculated from the generalized Callen-Suzuki identity \cite{call63,suzu65,balc02}:
\begin{eqnarray}
\langle S_j S_{j+1} \rangle = \langle S_{j+1} \tanh[\beta J (S_{j-1} + S_{j+1})] \rangle. 
\label{eq:iccor}
\end{eqnarray}  
Applying the differential operator and van der Waerden spin identity \cite{waer41} one still obtains an exact expression for the nearest-neighbor pair correlator:
\begin{eqnarray}
\langle S_j S_{j+1} \rangle = \frac{1}{2} (1 + \langle S_{j-1} S_{j+1} \rangle) \tanh (2 \beta J),
\label{eq:icor}
\end{eqnarray}
which is expressed in terms of the next-nearest-neighbor pair correlator $\langle S_{j-1} S_{j+1} \rangle$. To keep consistency with the standard formulation of the EFT \cite{kane93} one should however employ the decoupling approximation for the next-nearest-neighbor pair correlator $\langle S_{j-1} S_{j+1} \rangle \approx \langle S_{j-1} \rangle \langle S_{j+1} \rangle  = m^2$ in order to get the final formula for the nearest-neighbor pair correlator:
\begin{eqnarray}
\langle S_j S_{j+1} \rangle = \frac{1}{2} (1 + m^2) \tanh (2 \beta J) = \frac{1}{2} \tanh (2 \beta J). 
\label{eq:icorf}
\end{eqnarray}

\subsubsection{Exact results} 

The spin-1/2 Ising chain is perhaps the most famous lattice-statistical model, which can be exactly solved by several alternative approaches (see Ref. \cite{stre15} and references cited therein). The partition function of the spin-1/2 Ising chain can be cast within the transfer-matrix method \cite{kram41} to the following form:
\begin{eqnarray}
Z \!\!\!&=&\!\!\! \sum_{ \{S\} } \exp(-\beta H) = 
                  \sum_{ \{S\} } \prod_{j=1}^{N} \exp(\beta J S_j S_{j+1}) \nonumber \\
  \!\!\!&=&\!\!\! \sum_{ \{S\} } \prod_{j=1}^{N} {\rm T} \, (S_j, S_{j+1}).
\label{pfic}
\end{eqnarray}
The expression ${\rm T} \, (S_j, S_{j+1}) = \exp(\beta J S_j S_{j+1})$ can be repeatedly identified with two-by-two transfer matrix given by Eq. (\ref{tmd}), whereas the consecutive summation over spin states allows one to express the partition function in terms of a trace of $N$th power of the transfer matrix (\ref{tmd}):
\begin{eqnarray}
Z = \mbox{Tr}\, {\rm T}^N = \lambda_1^N + \lambda_2^N.
\label{pfictm}
\end{eqnarray}
After substituting two eigenvalues $\lambda_1 = 2 \cosh(\beta J)$ and $\lambda_2 = 2 \sinh(\beta J)$ of the transfer matrix (\ref{tmd}) into Eq. (\ref{pfictm}) one derives the following exact result for the partition function of the spin-1/2 Ising chain:
\begin{eqnarray}
Z = 2^N [\cosh^N (\beta J) + \sinh^N (\beta J)],
\label{pficf}
\end{eqnarray}
which holds regardless of whether the chain size is finite or not. The resultant expression (\ref{pficf}) for the partition function of the spin-1/2 Ising chain can be further simplified in the thermodynamic limit $N \to \infty$ to the following final form:
\begin{eqnarray}
Z = 2^N \cosh^N (\beta J).
\label{pficftl}
\end{eqnarray}

\begin{figure}[t]
\begin{center}
\includegraphics[width=0.9\columnwidth]{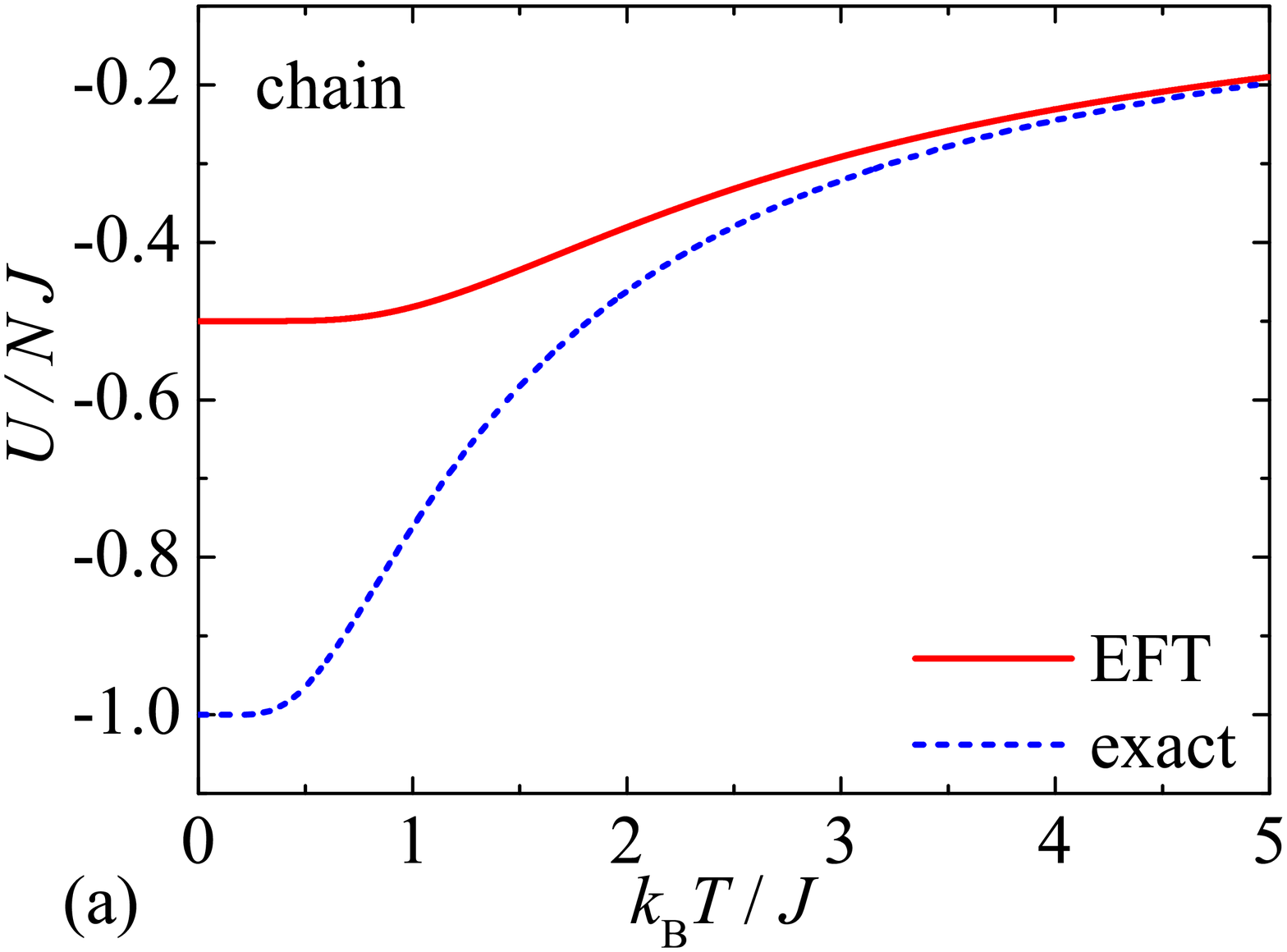}
\includegraphics[width=0.9\columnwidth]{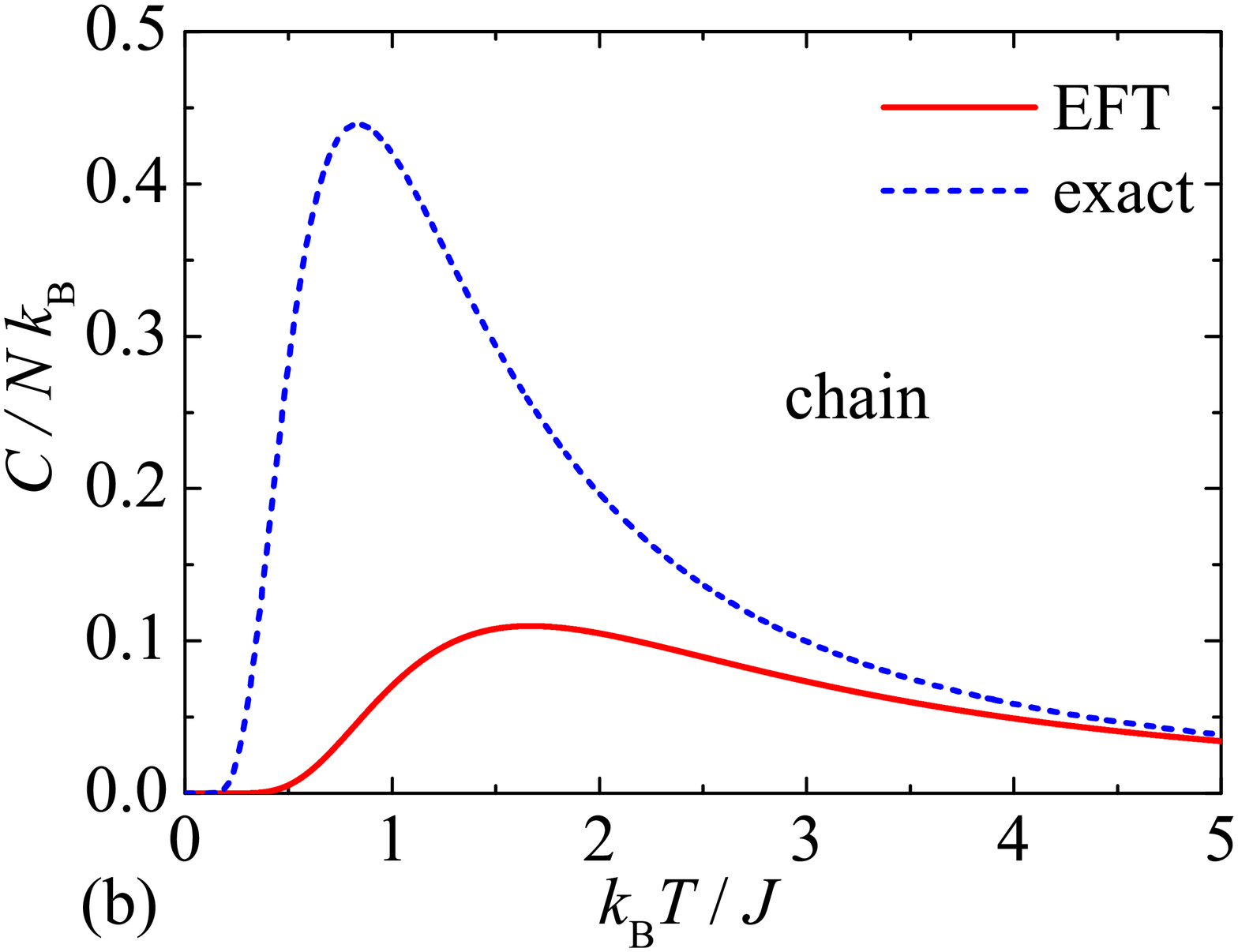}
\end{center}
\vspace*{-0.5cm}
\caption{Temperature variations of the internal energy (panel a) and specific heat (panel b) calculated for the spin-1/2 Ising chain within the EFT (solid lines) and exact method (broken lines).}
\label{figlcuc}
\end{figure}

It is evident from Eq. (\ref{pficftl}) that neither the partition function nor any of its temperature derivatives does not display singularity, which precludes existence of a spontaneous long-range order and a phase transition at nonzero temperatures. The exact result for the partition function (\ref{pficftl}) can be exploited in order to furnish evidence for vanishing spontaneous magnetization at any nonzero temperature:
\begin{eqnarray}
m \equiv \langle S_j \rangle = \frac{1}{Z} \sum_{\{ S\}} S_j \exp(-\beta H) = 0.
\label{micftl}
\end{eqnarray}
The internal energy of the spin-1/2 Ising chain can be also calculated from the partition function (\ref{pficftl}):
\begin{eqnarray}
U = -\frac{\partial \ln Z}{\partial \beta} =- N J \tanh (\beta J),
\label{uicftl}
\end{eqnarray}
which is consistent with the following exact expression for the nearest-neighbor pair correlator $\langle S_j S_{j+1} \rangle = \tanh (\beta J)$ that apparently differs from the one (\ref{eq:icorf}) derived within the EFT. The exact results for the internal energy and specific heat of the spin-1/2 Ising chain are compared in Fig. \ref{figlcuc} with the corresponding results of the EFT. In this particular case the EFT predicts smooth continuous temperature dependences precluding existence of the spontaneous long-range order and finite-temperature phase transition quite similarly as the exact transfer-matrix method does. Evidently, the results derived with the help of the EFT converge to the exact ones in a high-temperature region, while they start to significantly deviate from the exact ones in a low-temperature region due to presence of zero-temperature phase transition. It could be thus concluded that the standard EFT adopting the HK decoupling scheme for higher-order correlations \cite{kane93} does not capture exact results for all physical quantities though it correctly reproduces absence of the spontaneous magnetization and finite-temperature phase transition.  

\subsection{Two-leg ladder}
\label{sec:ladder}

The next example for our consideration is the spin-1/2 Ising two-leg ladder schematically shown in Fig. \ref{fig1d}(c) and mathematically defined through the Hamiltonian:
\begin{eqnarray}
H = - J \sum_{j=1}^N (S_{2j-1} S_{2j+1} + S_{2j} S_{2j+2}) - J_R \sum_{j=1}^N S_{2j-1} S_{2j},
\label{eq:hamlad}
\end{eqnarray}
which is composed of two coupled chains being considered under the periodic boundary condition $S_{2N+j} \equiv S_j$. The coupling constant $J$ denotes the intra-chain interaction, while the coupling constant $J_R$ refers to the inter-chain interaction along rungs of a two-leg ladder [see Fig. \ref{fig1d}(c)]. The solution for the spin-1/2 Ising two-leg ladder (\ref{eq:hamlad}) can be readily obtained within the EFT \cite{nl19a,nl19b,nl19c} as well as transfer-matrix method \cite{inou71,kalo75,fedr76,yoko89}, which will be briefly described in the following parts.  

\subsubsection{Effective-field theory} 

\begin{figure}[t]
\begin{center}
\includegraphics[width=0.9\columnwidth]{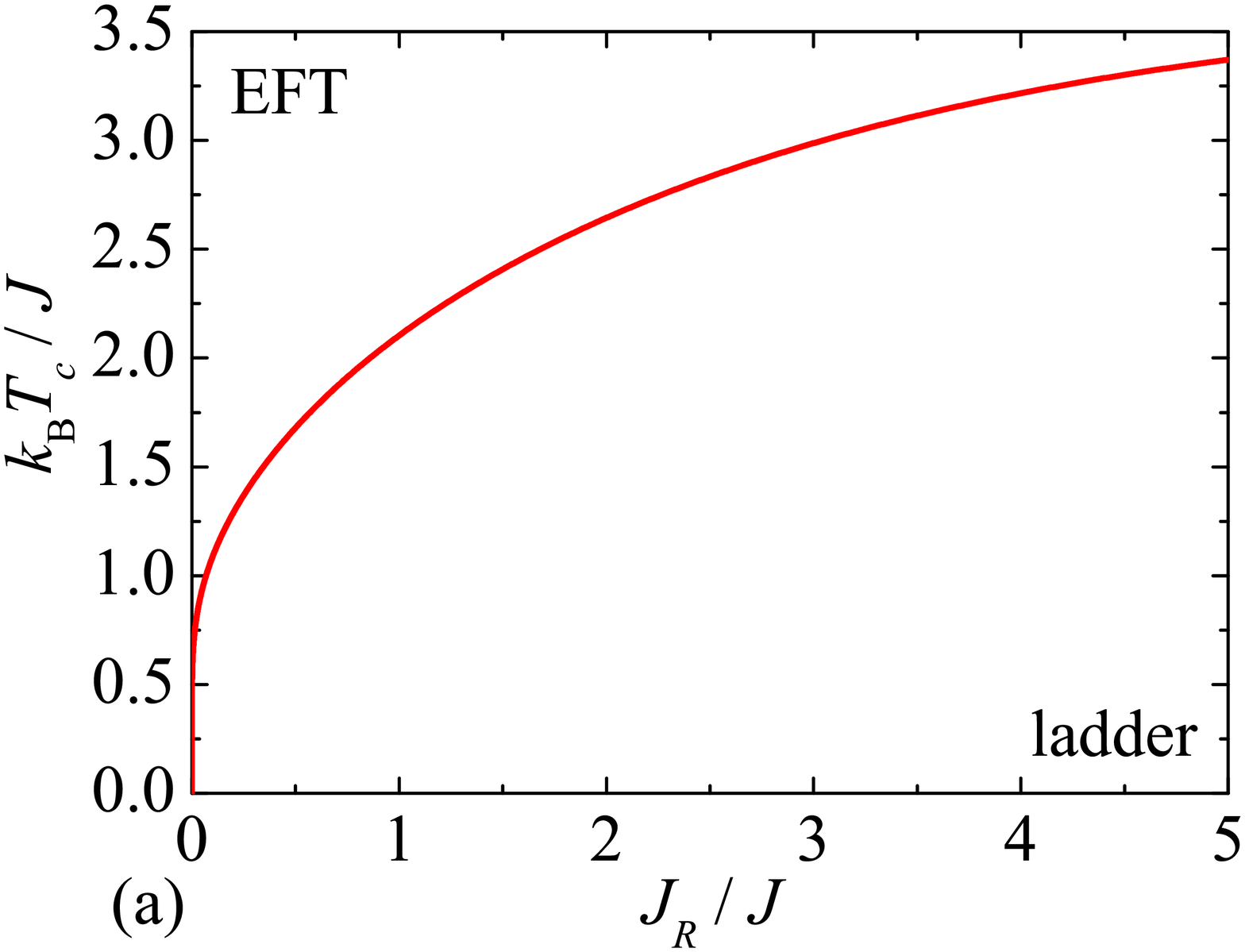}
\includegraphics[width=0.9\columnwidth]{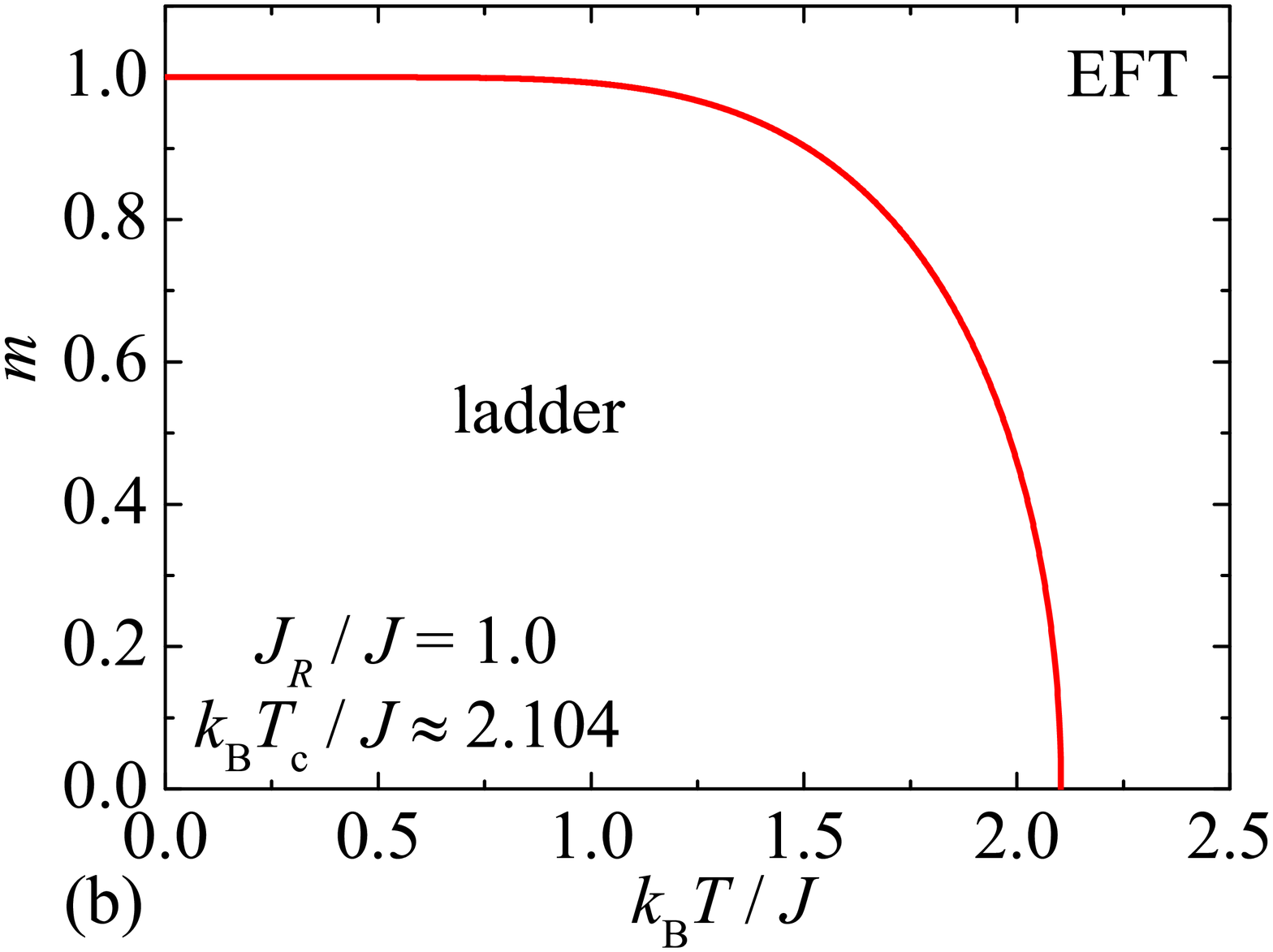}
\end{center}
\vspace*{-0.5cm}
\caption{(a) The critical temperature of the spin-1/2 Ising two-leg ladder calculated within the EFT as a function of the interaction ratio $J_R/J$; (b) Temperature variations of the spontaneous magnetization calculated for the spin-1/2 Ising ladder within the EFT for the particular value of the interaction ratio $J_R/J = 1$.}
\label{figladtcm}
\end{figure}

The local magnetization of the spin-1/2 Ising two-leg ladder given by the Hamiltonian (\ref{eq:hamlad}) can be calculated from the exact Callen-Suzuki spin identity \cite{call63,suzu65,balc02}:
\begin{eqnarray}
m \equiv \langle S_j \rangle = \langle \tanh[\beta J (S_{j-2} + S_{j+2}) + \beta J_R S_{j+1}] \rangle, 
\label{eq:lmag}
\end{eqnarray}  
which can be rewritten using the differential operator technique to the following form:
\begin{eqnarray}
m = \langle \exp[(S_{j-2} + S_{j+2}) \nabla_x + S_{j+1} \nabla_y] \rangle \tanh(\beta J x + \beta J_R y). 
\label{eq:lmdo}
\end{eqnarray}
Using the exact van der Waerden spin identity \cite{waer41}, the differential operator and the HK decoupling scheme \cite{kane93} for higher-order correlations  enables one to derive the following result for the magnetization: 
\begin{eqnarray}
m = m (2 L_1 + L_2) + m^3 L_3, 
\label{eq:lmvw}
\end{eqnarray} 
which is expressed in terms of the coefficients $L_1-L_3$ given by Eq. (\ref{eq:dc}). According to Eq. (\ref{eq:lmvw}), the magnetization of the spin-1/2 Ising ladder can be calculated from the formula:
\begin{eqnarray}
m = \sqrt{\frac{1 - 2L_1 - L_2}{L_3}}, 
\label{eq:lmf}
\end{eqnarray}
which gives evidence for a nonzero spontaneous magnetization below the critical temperature given by the critical condition $2 L_1 + L_2 = 1$. The critical temperature calculated from this critical condition is displayed in Fig. \ref{figladtcm}(a) as a function of the interaction ratio $J_R/J$ and this phase boundary is in a perfect agreement with the one reported in Fig. 2 of Ref. \cite{nl19a}. It follows from this figure that the critical temperature monotonically increases with increasing of the interaction ratio $J_R/J$, whereas it exhibits a sudden drop down to zero as the interaction ratio vanishes $J_R/J \to 0$. The spin-1/2 Ising two-leg ladder decomposes in the limiting case $J_R/J = 0$ into two separate chains involving only spins with the coordination number two and hence, the EFT only predicts a spontaneous long-range order just for the spin-1/2 Ising two-leg ladder with $J_R/J \neq 0$ involving spins with the coordination number three. Temperature dependence of the spontaneous magnetization of the spin-1/2 Ising ladder presented in Fig. \ref{figladtcm}(b) for one illustrative case of the interaction ratio $J_R/J = 1$ convincingly evidences a spontaneous long-range order, which disappears above the critical temperature $k_B T_c/J \approx 2.104$. 

The internal energy of the spin-1/2 Ising ladder can be expressed in terms of two nearest-neighbor pair correlators $U = -2NJ \langle S_{j} S_{j+2} \rangle - NJ_R \langle S_{2j-1} S_{2j} \rangle$, which can be  evaluated with the help of the generalized Callen-Suzuki identities \cite{call63,suzu65,balc02}:
\begin{eqnarray}
\langle S_j S_{j+2} \rangle \!\!\!&=&\!\!\! \langle S_{j+2} \tanh[\beta J (S_{j-2} + S_{j+2}) + \beta J_R S_{j+1}] \rangle,  \nonumber \\
\langle S_{2j-1} S_{2j} \rangle \!\!\!&=&\!\!\! \langle S_{2j} \tanh[\beta J (S_{2j-3} + S_{2j+1}) + \beta J_R S_{2j}] \rangle. 
\label{eq:lcor}
\end{eqnarray}  
By making use of the differential operator, van der Waerden spin identity \cite{waer41} and the HK decoupling approximation \cite{kane93} introduced within the standard formulation of the EFT  one obtains the following final formulas for the nearest-neighbor pair correlators:
\begin{eqnarray}
\langle S_j S_{j+2} \rangle \!\!\!&=&\!\!\! L_1 + m^2 (L_1 + L_2 + L_3),  \nonumber \\
\langle S_{2j-1} S_{2j} \rangle \!\!\!&=&\!\!\! L_2 + m^2 (2L_1 + L_3). 
\label{eq:lcorf}
\end{eqnarray}

\subsubsection{Exact results} 

\begin{figure}[t]
\begin{center}
\includegraphics[width=0.9\columnwidth]{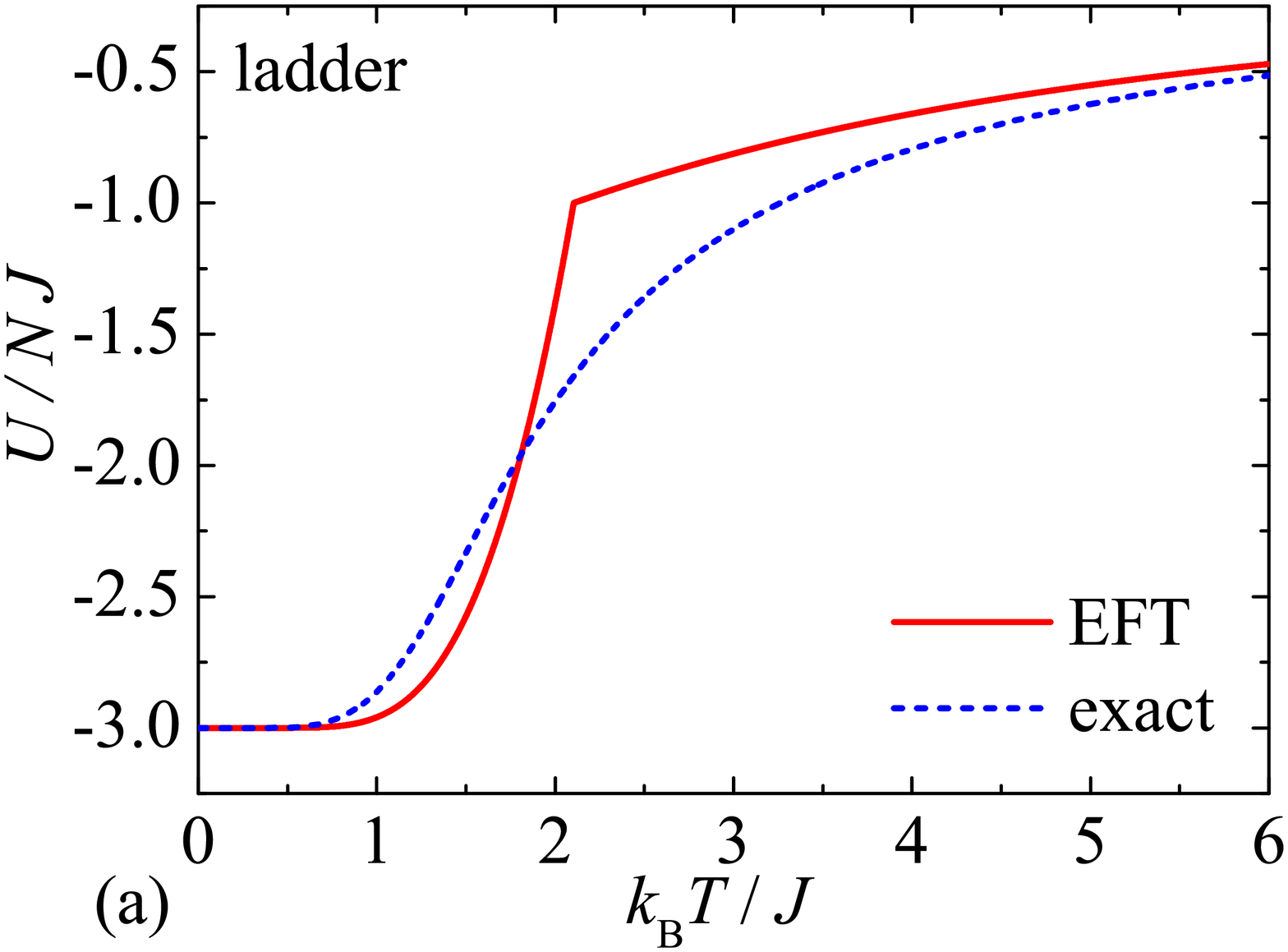}
\includegraphics[width=0.9\columnwidth]{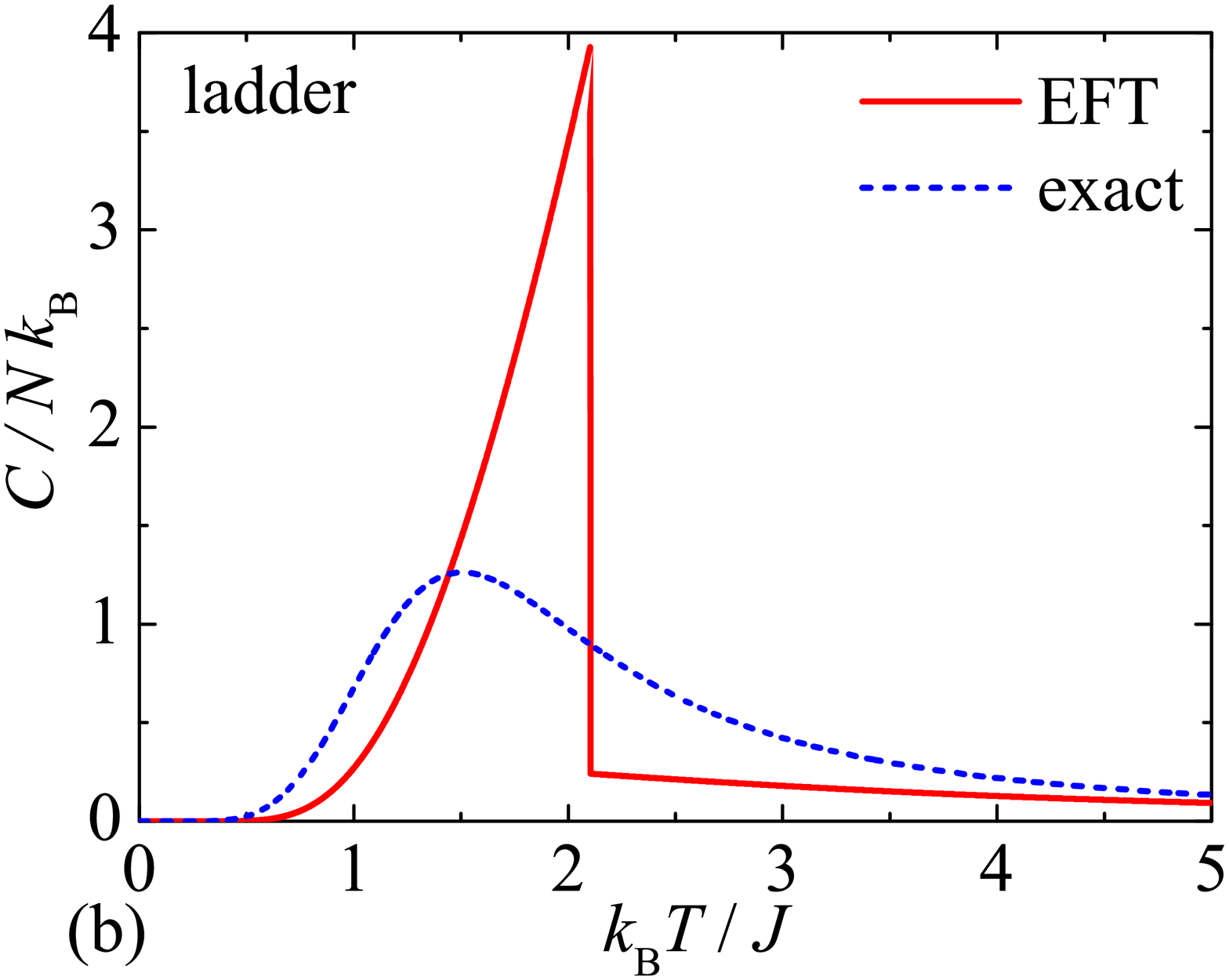}
\end{center}
\vspace*{-0.5cm}
\caption{Temperature variations of the internal energy (panel a) and specific heat (panel b) calculated for the spin-1/2 Ising two-leg ladder with the specific value of the interaction ration $J_R/J = 1$ within the EFT (solid lines) and exact method (broken lines).}
\label{figladuc}
\end{figure}

The spin-1/2 Ising two-leg ladder has been exactly solved by exploiting the transfer-matrix approach in Refs. \cite{inou71,kalo75,fedr76,yoko89}, so we will recall here only a few most important steps of this derivation. Within this technique the partition function of the spin-1/2 Ising ladder can be written in the following form:
\begin{eqnarray}
Z \!\!\!&=&\!\!\! \sum_{ \{S\} } \prod_{j=1}^{N} \exp\left[\beta J (S_{2j-1} S_{2j+1} + S_{2j} S_{2j+2}) \right. \nonumber \\
 \!\!\!&+&\!\!\! \left. \frac{\beta J_R}{2} (S_{2j-1} S_{2j} + S_{2j+1} S_{2j+2}) \right] \nonumber \\
  \!\!\!&=&\!\!\! \sum_{ \{S\} } \prod_{j=1}^{N} {\rm T} \, (S_{2j-1}, S_{2j}; S_{2j+1}, S_{2j+2}).
\label{pfil}
\end{eqnarray}
The expression ${\rm T} \, (S_{2j-1}, S_{2j}; S_{2j+1}, S_{2j+2}) =$ $\exp[\beta J (S_{2j-1} S_{2j+1} + S_{2j} S_{2j+2}) + \frac{\beta J_R}{2} (S_{2j-1} S_{2j} + S_{2j+1} S_{2j+2})]$ can be identified as four-by-four transfer matrix:
\begin{eqnarray}
{\rm T} \!\!\!\!\!\!&(&\!\!\!\!\!\!S_{2j-1}, S_{2j}; S_{2j+1}, S_{2j+2}) \nonumber \\
\!\!\!&=&\!\!\! 
\left( \!\begin{array}{cccc}
       \! {\rm e}^{2\beta J + \beta J_R} & 1 & 1 & {\rm e}^{-2\beta J + \beta J_R} \\ 
			 \! 1 & {\rm e}^{2\beta J - \beta J_R} & {\rm e}^{-2\beta J - \beta J_R} & 1 \\
			 \! 1 & {\rm e}^{-2\beta J - \beta J_R} & {\rm e}^{2\beta J - \beta J_R} & 1 \\
			 \! {\rm e}^{-2\beta J + \beta J_R} & 1 & 1 & {\rm e}^{2\beta J + \beta J_R} 								 
                          \end{array} 
                  \! \right)\!.
\label{tml}
\end{eqnarray}
The consecutive summation over spin states in Eq. (\ref{pfil}) leads to the trace of $N$th power of the transfer matrix (\ref{tml}):
\begin{eqnarray}
Z = \mbox{Tr}\, {\rm T} \,^N = \lambda_1^N + \lambda_2^N + \lambda_3^N + \lambda_4^N,
\label{pfiltm}
\end{eqnarray}
which is expressed in terms of four transfer-matrix eigenvalues: 
\begin{eqnarray}
\lambda_1 \!\!\!&=&\!\!\! 2 \exp(\beta J_R) \sinh(2 \beta J), \nonumber\\
\lambda_2 \!\!\!&=&\!\!\! 2 \exp(-\beta J_R) \sinh(2 \beta J), \nonumber\\
\lambda_{3,4} \!\!\!&=&\!\!\! 2 [\cosh(2 \beta J) \cosh(\beta J_R) \nonumber\\
\!\!\!&\pm&\!\!\! \sqrt{\cosh^2(2 \beta J) \sinh^2(\beta J_R) + 1}]. 
\label{iltme}
\end{eqnarray}
As usual, the partition function is governed within the transfer-matrix method in the thermodynamic limit $N \to \infty$ by the largest transfer-matrix eigenvalue, whereas the final formula for the partition function can be written in this compact form:
\begin{eqnarray}
Z \!\!\!&=&\!\!\! 2^N [\cosh(2 \beta J) \cosh(\beta J_R) \nonumber \\
  \!\!\!&+&\!\!\! \sqrt{\cosh^2(2 \beta J) \sinh^2(\beta J_R) + 1}]^N.
\label{pflcftl}
\end{eqnarray}
It can be easily proved that the partition function (\ref{pflcftl}) and all its temperature derivatives are free of any mathematical singularities, which rules out presence of a spontaneous long-range order and finite-temperature phase transition. The exact result for the partition function (\ref{pflcftl}) can be also employed in order to evidence a null spontaneous magnetization at any nonzero temperature. The internal energy and specific heat of the spin-1/2 Ising two-leg ladder can be calculated from the partition function (\ref{pflcftl}) by standard means $U = -\frac{\partial \ln Z}{\partial \beta}$ and $C = \frac{\partial U}{\partial T}$. The as-obtained exact results for the internal energy and specific heat of the spin-1/2 Ising two-leg ladder are shown in Fig. \ref{figladuc} together with the corresponding results of the EFT. It is clear that the results stemming from the EFT coincide with the exact results in low- and high-temperature regions, while they significantly deviate from the exact ones in a neighborhood of the critical temperature where the internal energy and specific heat exhibit a cusp and finite jump, respectively. Bearing all this in mind, the standard EFT adopting the HK decoupling scheme \cite{kane93} for higher-order correlations fails in predicting absence of the spontaneous long-range order and phase transition of this one-dimensional spin system. 

\subsection{Hexagonal ladder}
\label{sec:hexladder}

The last paradigmatic example of one-dimensional spin systems we are going to consider is the spin-1/2 Ising hexagonal ladder defined through the Hamiltonian:
\begin{eqnarray}
H = - J \sum_{j=1}^{2N} \sigma_{j} (S_{j} + S_{j+2}) - J_R \sum_{j=1}^N S_{2j-1} S_{2j}.
\label{eq:hamhlad}
\end{eqnarray}
The spin-1/2 Ising hexagonal ladder given by Eq. (\ref{eq:hamhlad}) refers to two coupled zig-zag chains considered under the periodic boundary condition $S_{2N+j} \equiv S_j$  [see Fig. \ref{fig1d}(d)]. The coupling constant $J$ denotes the intra-chain interaction within both zig-zag chains, while the coupling constant $J_R$ denotes the inter-chain interaction that couples together two zig-zag chains of a hexagonal ladder. The  spin-1/2 Ising hexagonal ladder (\ref{eq:hamhlad}) can be easily solved with the help of EFT \cite{nl19c} and exact method, which will be briefly described hereafter.  

\subsubsection{Effective-field theory} 

\begin{figure}[t]
\begin{center}
\includegraphics[width=0.9\columnwidth]{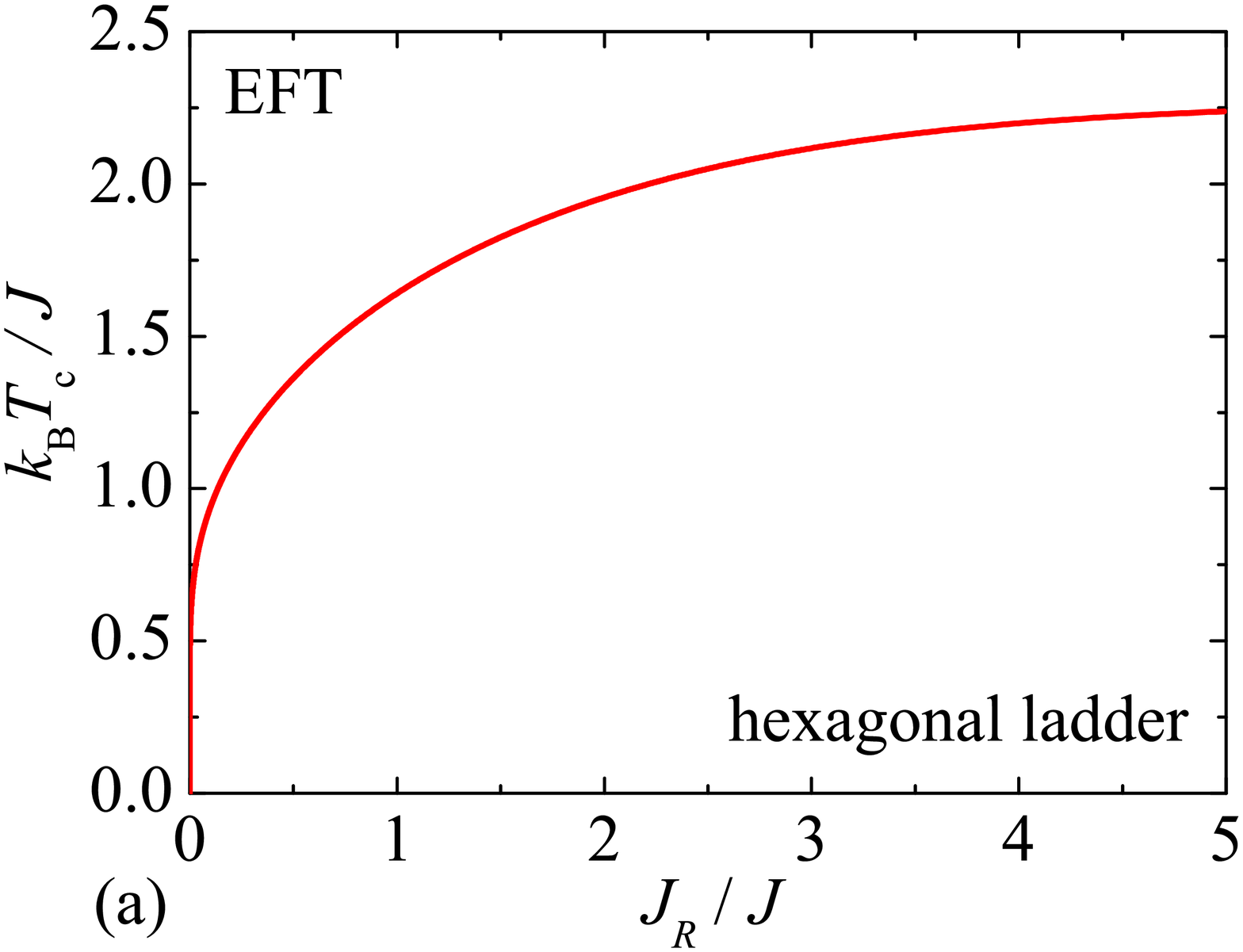}
\includegraphics[width=0.9\columnwidth]{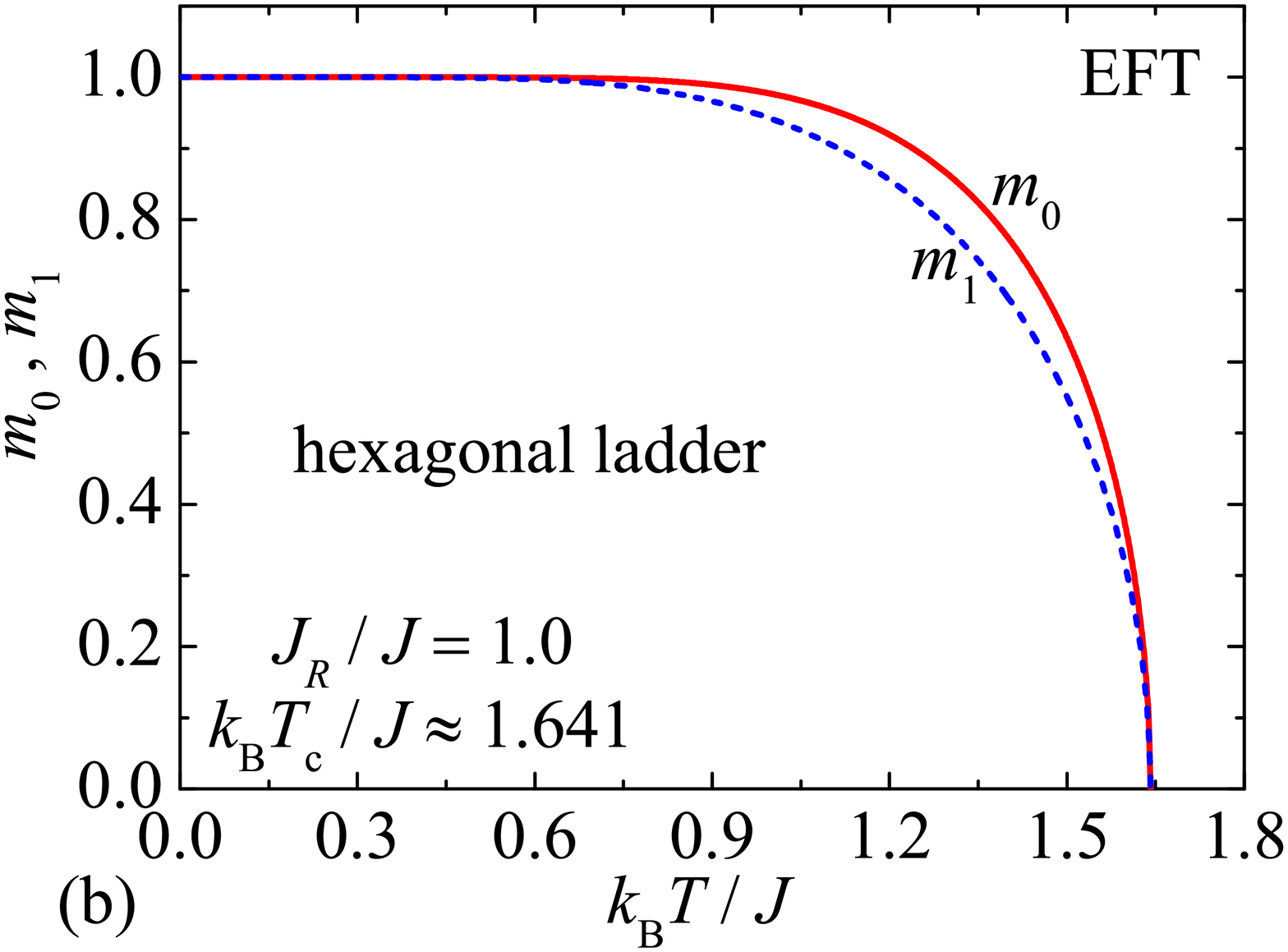}
\end{center}
\vspace*{-0.5cm}
\caption{(a) The critical temperature of the spin-1/2 Ising hexagonal ladder calculated within the EFT as a function of the interaction ratio $J_R/J$; (b) Temperature variations of the spontaneous magnetization calculated for the spin-1/2 Ising hexagonal ladder within the EFT for the particular value of the interaction ratio $J_R/J = 1$.}
\label{fighltcm}
\end{figure}

The local magnetizations of the spin-1/2 Ising hexagonal ladder can be derived from the exact Callen-Suzuki spin identities \cite{call63,suzu65,balc02}:
\begin{eqnarray}
m_0 \!\!\!&\equiv&\!\!\! \langle S_{j} \rangle = \langle \tanh[\beta J (\sigma_{j-2} + \sigma_{j}) 
+ \beta J_R S_{j \pm 1}] \rangle,  \nonumber \\
m_1 \!\!\!&\equiv&\!\!\! \langle \sigma_j \rangle = \langle \tanh[\beta J (S_{j} + S_{j+2})] \rangle, 
\label{eq:hlmag}
\end{eqnarray}  
whereas the plus (minus) sign in the first equation refers to odd (even) spins $S_{2j-1}$ ($S_{2j}$). The Callen-Suzuki spin identities (\ref{eq:hlmag}) can be recast with the help of differential operator technique into the following form:
\begin{eqnarray}
m_0 \!\!\!&=&\!\!\! \langle \exp[(\sigma_{j-2} \!+\! \sigma_{j}) \nabla_x \!+\! S_{j \pm 1} \nabla_y] \rangle 
\tanh(\beta J x \!+\! \beta J_R y), \nonumber \\
m_1 \!\!\!&=&\!\!\! \langle \exp[(S_{j} \!+\! S_{j+2}) \nabla_x] \rangle \tanh(\beta J x).
\label{eq:hlmdo}
\end{eqnarray}
By combining the exact van der Waerden spin identity \cite{waer41} with the differential operator and the HK decoupling scheme \cite{kane93} for higher-order correlations one gets the following couple of equations for the local magnetizations: 
\begin{eqnarray}
m_0 \!\!\!&=&\!\!\! 2 m_1 L_1 + m_0 L_2 + m_0 m_1^2 L_3, \nonumber \\
m_1 \!\!\!&=&\!\!\! m_0 L_4,
\label{eq:hlmvw}
\end{eqnarray} 
which are expressed in terms of the coefficients $L_1-L_3$ given by Eq. (\ref{eq:dc}) and the newly defined coefficient $L_4= \tanh(2 \beta J)$. Eliminating either the expression $m_0$ or $m_1$ from the couple of equations (\ref{eq:hlmvw}) one obtains the following final formulas for the local magnetizations of the spin-1/2 Ising hexagonal ladder:
\begin{eqnarray}
m_0 \!\!\!&=&\!\!\! \sqrt{\frac{1 - 2 L_1 L_4 - L_2}{L_3 L_4^2}}, \nonumber \\
m_1 \!\!\!&=&\!\!\! \sqrt{\frac{1 - 2 L_1 L_4 - L_2}{L_3}}.
\label{eq:hlmf}
\end{eqnarray}
The above formulas are consistent with presence of a nonzero spontaneous magnetization below the critical temperature given by the critical constraint $2 L_1 L_4 + L_2 = 1$. The critical temperature calculated according to this critical condition is plotted in Fig. \ref{fighltcm}(a) depending on the interaction ratio $J_R/J$. Evidently, the critical temperature rises steadily with increasing of the interaction ratio $J_R/J$ and it shows a sudden drop down to zero as the interaction ratio vanishes $J_R/J \to 0$. The spin-1/2 Ising hexagonal ladder decomposes in the particular limit $J_R/J = 0$ into two separate zig-zag chains involving only spins with the coordination number two and thus, the EFT predicts a spontaneous long-range order just for the spin-1/2 Ising hexagonal ladder with $J_R/J \neq 0$ involving spins with the coordination number three. To support this statement, temperature variations of the spontaneous magnetization of the spin-1/2 Ising hexagonal ladder are depicted in Fig. \ref{fighltcm}(b) for the special value of the interaction ratio $J_R/J = 1$, which convincingly evidences presence of a spontaneous long-range order below the critical temperature $k_B T_c/J \approx 1.641$. 

The internal energy of the spin-1/2 Ising hexagonal ladder can be evaluated from two nearest-neighbor pair correlators $U = -4NJ \langle S_{j} \sigma_{j} \rangle - NJ_R \langle S_{2j-1} S_{2j} \rangle$, which can be also calculated using the generalized Callen-Suzuki identities \cite{call63,suzu65,balc02}:
\begin{eqnarray}
\langle S_j \sigma_{j} \rangle \!\!\!&=&\!\!\! \langle \sigma_{j} \tanh[\beta J (\sigma_{j-2} + \sigma_{j}) 
+ \beta J_R S_{j \pm 1}] \rangle,  \nonumber \\
\langle S_{2j-1} S_{2j} \rangle \!\!\!&=&\!\!\! \langle S_{2j} \tanh[\beta J (\sigma_{2j-3} + \sigma_{2j+1}) 
+ \beta J_R S_{2j}] \rangle. 
\label{eq:hlcor}
\end{eqnarray}  
By taking advantage of the differential operator, van der Waerden spin identity \cite{waer41} and the HK decoupling approximation \cite{kane93} within the standard formulation of the EFT  one gets the following final formulas for the nearest-neighbor pair correlators:
\begin{eqnarray}
\langle S_j \sigma_{j} \rangle  \!\!\!&=&\!\!\! (1 + m_1^2) L_1 + m_0 m_1 (L_2 + L_3),  \nonumber \\
\langle S_{2j-1} S_{2j} \rangle \!\!\!&=&\!\!\! 2 m_0 m_1 L_1 + L_2 + m_1^2 L_3. 
\label{eq:hlcorf}
\end{eqnarray}

\subsubsection{Exact results} 

The spin-1/2 Ising hexagonal ladder given by the Hamiltonian (\ref{eq:hamhlad}) can be alternatively viewed as a two-leg ladder formed by 'nodal' spins $S_j$, the horizontal bonds of which contain additional 'decorating' spins $\sigma_j$. The exact solution for the spin-1/2 Ising hexagonal ladder can be accordingly achieved by combining the decoration-iteration transformation \cite{syoz51,fish59} with the transfer-matrix approach \cite{kram41}. To this end, let us rewrite first the partition function of the spin-1/2 Ising hexagonal ladder to the following form:
\begin{eqnarray}
Z = \sum_{ \{S\} } \prod_{j=1}^{N} \!\!\!\!\!&&\!\!\!\!\!
     \exp\left[\frac{\beta J_R}{2} (S_{2j-1} S_{2j} + S_{2j+1} S_{2j+2}) \right] \nonumber \\ \!\!\!\!\!&\times&\!\!\!\!\!		  \sum_{\sigma_{2j-1} = \pm 1} \exp[\beta J \sigma_{2j-1} (S_{2j-1} + S_{2j+1})] 
	\nonumber \\ \!\!\!\!\!&\times&\!\!\!\!\!
  \sum_{\sigma_{2j} = \pm 1} \exp[\beta J \sigma_{2j} (S_{2j} + S_{2j+2})].
\label{pfihl}
\end{eqnarray}
Here, we have used the fact that the summation over states of the 'decorating' spins $\sigma_j$ can be performed independently of each other and before performing summation $\sum_{ \{S\} }$ over states of all 'nodal' spins $S_j$. The Boltzmann's weights given in the second and third line of Eq. (\ref{pfihl}) can be substituted by a simpler expression through the so-called decoration-iteration transformation \cite{syoz51,fish59}:
\begin{eqnarray}
\sum_{\sigma_{j} = \pm 1} \exp[\beta J \sigma_{j} (S_{j} \!\!\!\!\!&+&\!\!\!\!\! S_{j+2})] =2 \cosh [\beta J (S_{j} + S_{j+2})] \nonumber \\ \!\!\!\!\!&=&\!\!\!\!\! A \exp[\beta J_{eff} (S_{j} + S_{j+2})],
\label{eq:dit}
\end{eqnarray}
which should hold for all four possible states of two nodal spins $S_j$ and $S_{j+2}$ in order to ensure general validity of this mapping transformation. This 'self-consistency' condition unambiguously determines the mapping parameters $A$ and $\beta J_{eff}$ introduced in the decoration-iteration transformation (\ref{eq:dit}):
\begin{eqnarray}
A \!\!\!&=&\!\!\! 2 \sqrt{\cosh(2 \beta J)}, \nonumber \\
\beta J_{eff} \!\!\!&=&\!\!\! \frac{1}{2} \ln [\cosh(2 \beta J)].
\label{eq:aj}
\end{eqnarray}
Substituting the decoration-iteration transformation (\ref{eq:dit}) into Eq. (\ref{pfihl}) provides the following expression for the partition function:
\begin{eqnarray}
Z \!\!\!&=&\!\!\! A^{2N}  \sum_{ \{S\} } \prod_{j=1}^{N} 
\exp\left[\beta J_{eff} (S_{2j-1} S_{2j+1} + S_{2j} S_{2j+2}) \right. \nonumber \\
 \!\!\!&+&\!\!\! \left. \frac{\beta J_R}{2} (S_{2j-1} S_{2j} + S_{2j+1} S_{2j+2}) \right].
\label{eq:pfhld}
\end{eqnarray}
Apart from the trivial prefactor $A^{2N}$, the partition function (\ref{eq:pfhld}) of the spin-1/2 Ising hexagonal ladder becomes formally equivalent with the expression (\ref{pfil}) determining the partition function of the spin-1/2 Ising two-leg ladder whenever the intra-chain interaction $J$ along legs of a two-leg ladder is replaced with the mapping parameter $J_{eff}$ given by Eq. (\ref{eq:aj}). This fact proves a rigorous mapping correspondence, which connects the partition function $Z$ of the spin-1/2 Ising hexagonal ladder with the intra- and inter-chain coupling constants $J$ and $J_R$ with the partition function $Z_{eff}$ of the effective spin-1/2 Ising two-leg ladder with the intra- and inter-chain coupling constants $J_{eff}$ and $J_R$:
\begin{eqnarray}
Z (\beta, J, J_R) = A^{2N}  Z_{eff} (\beta, J_{eff}, J_R).
\label{eq:pfhldit}
\end{eqnarray}
The exact result for the partition function of the spin-1/2 Ising hexagonal ladder can be thus obtained from the mapping relationship (\ref{eq:pfhldit}) by employing formerly derived exact result (\ref{pflcftl}) for the partition function of the spin-1/2 Ising two-leg ladder. After some algebra one obtains the following final result for the partition function of the spin-1/2 Ising hexagonal ladder:
\begin{eqnarray}
Z \!\!\!&=&\!\!\! 8^N \cosh^N (2 \beta J) [\cosh(2 \beta J_{eff}) \cosh(\beta J_R) \nonumber \\
  \!\!\!&+&\!\!\! \sqrt{\cosh^2(2 \beta J_{eff}) \sinh^2(\beta J_R) + 1}]^N,
\label{zhljef}
\end{eqnarray}
which is expressed in terms of the effective interaction $\beta J_{eff} = \frac{1}{2} \ln [\cosh(2 \beta J)]$. The partition function (\ref{zhljef}) and all its temperature derivatives are again completely free of mathematical singularities, which evidences that a nonzero critical temperature and spontaneous magnetization in Fig. \ref{fighltcm} are again just artifacts of the used approximate EFT. 

\begin{figure}[t]
\begin{center}
\includegraphics[width=0.9\columnwidth]{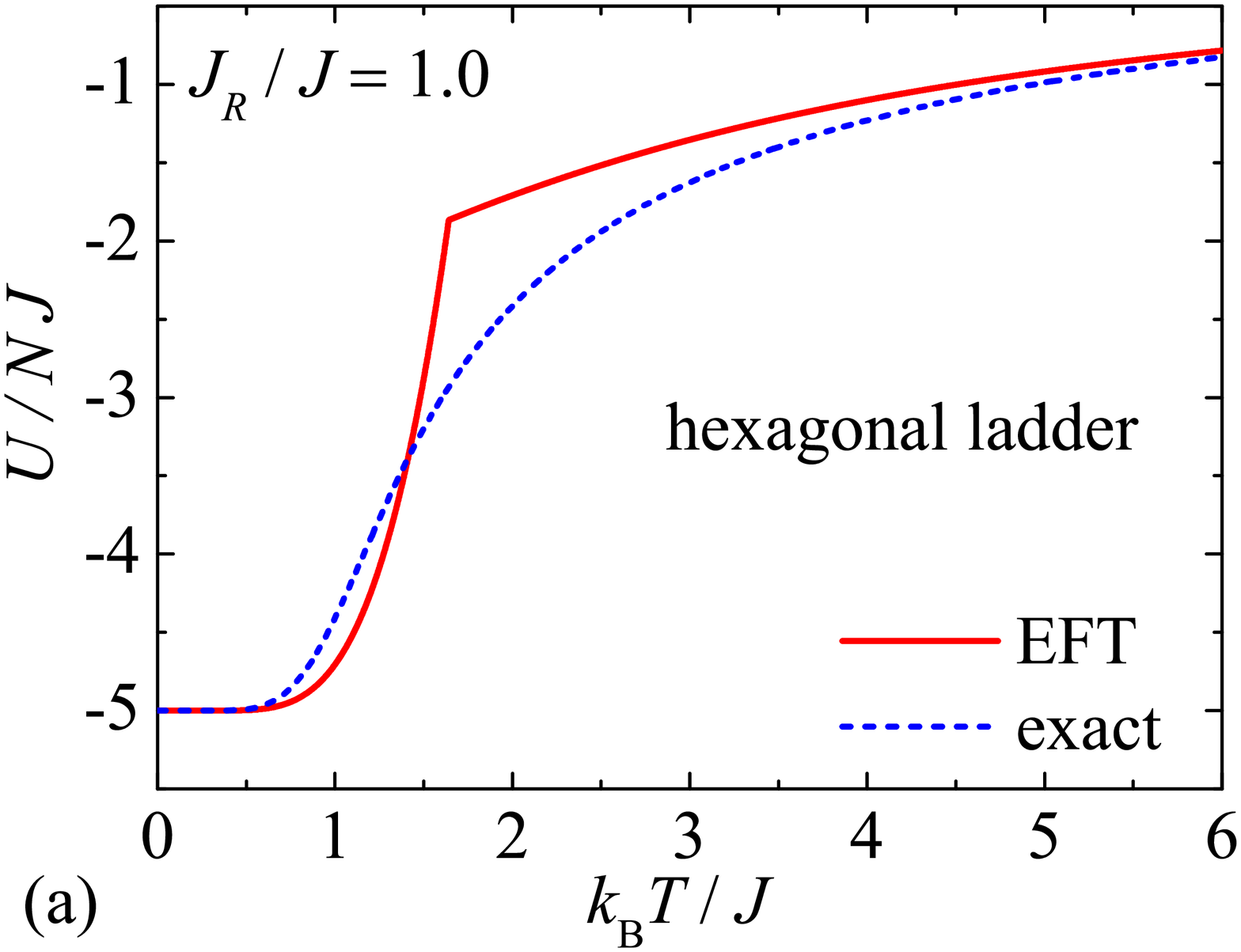}
\includegraphics[width=0.9\columnwidth]{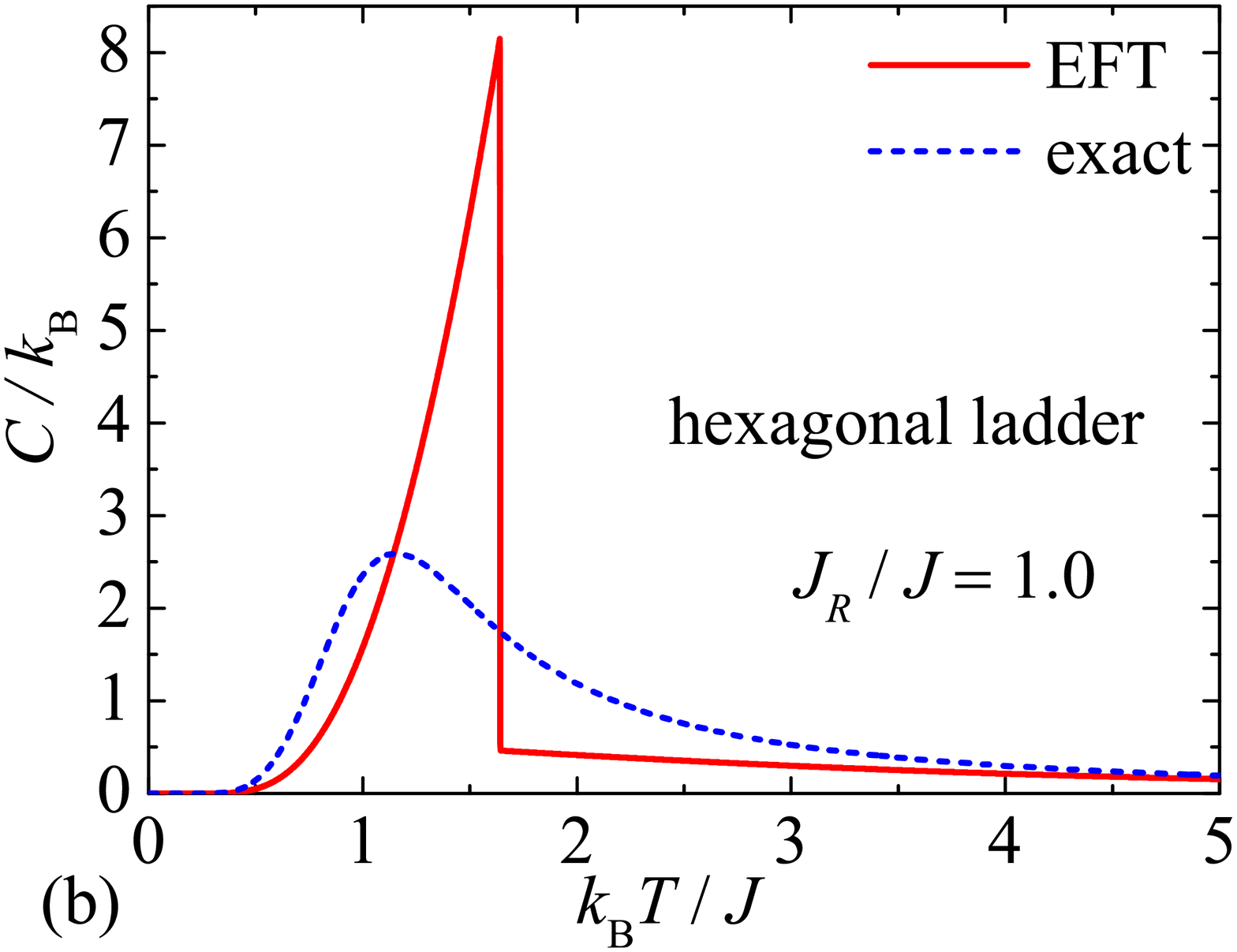}
\end{center}
\vspace*{-0.5cm}
\caption{Temperature variations of the internal energy (panel a) and specific heat (panel b) calculated for the spin-1/2 Ising hexagonal ladder with the specific value of the interaction ration $J_R/J = 1$ within the EFT (solid lines) and exact method (broken lines).}
\label{fighluc}
\end{figure}

The internal energy and specific heat of the spin-1/2 Ising hexagonal ladder can be easily calculated from the partition function (\ref{zhljef}) using the relations $U = -\frac{\partial \ln Z}{\partial \beta}$ and $C = \frac{\partial U}{\partial T}$. The exact results for the internal energy and specific heat of the spin-1/2 Ising hexagonal ladder are displayed in Fig. \ref{fighluc} along with the corresponding results of the EFT. It is obvious that the exact results coincide with the ones derived with the help of the EFT in low- and high-temperature regions, while the largest discrepancy can be detected at moderate temperatures where the internal energy and specific heat exhibit according to the EFT a cusp and a finite jump in opposite with smooth continuous changes acquired from the exact calculation. It could be thus concluded that the standard EFT adopting the decoupling scheme for higher-order correlations \cite{kane93} fails in predicting absence of the spontaneous long-range order and phase transition of this one-dimensional spin system.

\section{On the non-existence of phase transition and the failure of effective-field theory: general remarks}
\label{sec:4}

Exact solutions presented in two previous sections for a few selected examples of zero- and one-dimensional Ising spin systems clearly demonstrate non-existence of phase transition at any finite temperature. It is therefore quite plausible to address the question whether absence of phase transition is general feature and one may refute presence of a spontaneous long-range order for all zero- and one-dimensional Ising spin systems. Before doing this, two important comments are in order as far as a magnetic dimension of Ising spin systems is concerned, because this term is also often common cause of confusion. The Ising spin system is said to be zero-dimensional if it consists of a finite number of spins $N$ and this definition holds true regardless of its spatial geometry, which may be even higher-dimensional in a real space. The term zero-dimensional Ising spin systems thus refers to finite-size spin clusters as for instance the star with $N=4$, the cube with $N=8$, the star of David and decorated hexagonal nanoparticle with $N=12$ schematically shown in Fig. \ref{fig0d}. On the other hand, the Ising spin system is said to be one-dimensional if it involves infinite number of interacting spins in one spatial direction and either zero or finite number of interacting spins in other two spatial directions. Note furthermore that the infinite number of interacting spins along one spatial direction does not need to be necessarily placed on a single straight line. The previously studied examples of a branched chain, a sawtooth ($\Delta$) chain, two-leg and hexagonal ladders schematically shown in Fig. \ref{fig1d} accordingly belong to one-dimensional Ising spin systems quite similarly as a trivial example of a linear chain.   

The Ising model generally exhibits a phase transition if and only if one detects a non-analytic point in the respective density of free energy:
\begin{eqnarray}
F = - \frac{1}{N} k_B T \ln Z (N, T).
\label{eq:fed}
\end{eqnarray}
The non-analyticity means that the free-energy density (\ref{eq:fed}) cannot be expanded into the Taylor series in a vicinity of the critical temperature $T_c$, whereas this singular point is usually accompanied with a discontinuity or a divergence of some of its temperature derivatives though there are a few rare exceptions to this rule. It is immediately clear from Eq. (\ref{eq:fed}) that the free-energy density of zero-dimensional Ising spin systems cannot exhibit any non-analytic point, because the partition function $Z = \sum_{\{S\}} \exp(-\beta H)$ is just simple sum of a finite number of positive terms with character of smooth, continuous and differentiable exponential functions (Boltzmann's weights). All zero-dimensional Ising spin systems are accordingly free of any finite-temperature phase transition. It could be thus concluded that the spontaneous long-range order and the associated criticality  previously reported for several zero-dimensional Ising spin systems, which were mostly referred to as  nanoparticles \cite{np05,np09,np12a,np12b,np13a,np13b,np14a,np14b,np15a,np15b,np15c,np15d,np15e,np15f,np15g,np16a,np16b,np16c,np16d,np17a,np17b,np17c,np18a,np18b,np19a,np19b,np19c,np19d,np19e,np19f,np19g,np19h,np20a,np20b,np20c,np20d,np20e,np20f,np20g,np20h,np20i,np20j,np20k} or nanoislands \cite{ni13,ni15a,ni15b,ni15c,ni15d,ni15e,ni19a,ni19b,ni19c}, are just consequences of the failure originating from the HK decoupling approximation within the standard formulation of EFT \cite{kane93}. 

Compared to this, it is much more intricate to refute existence of a phase transition in one-dimensional Ising spin systems with infinite number of spins. Consequently, the spurious phase transition should be related to some non-analytic (singular) point of the free-energy density calculated in the thermodynamic limit $N \to \infty$:  
\begin{eqnarray}
F = - \lim_{N \to \infty} \frac{1}{N} k_B T \ln Z (N, T).
\label{eq:fedo}
\end{eqnarray}
The absence of non-analytic point in the free-energy density (\ref{eq:fedo}) cannot be definitely ruled out with the help of argument used previously for zero-dimensional Ising spin systems, because the partition function $Z = \sum_{\{S\}} \exp(-\beta H)$ is composed of infinite number of positive Boltzmann's weights that could eventually give rise to a mathematical singularity. However, the exact solution of any one-dimensional Ising spin system can be formulated within the transfer-matrix method, which allows one to express the partition function in terms of the largest transfer-matrix eigenvalue \cite{kram41}. According to the the Perron-Frobenius theorem \cite{meye00}, the largest eigenvalue of a positive finite transfer matrix is non-degenerate and one may thus simply refute a possible existence of a discontinuous phase transition. Moreover, the largest eigenvalue of a positive finite transfer matrix of one-dimensional Ising spin systems is simultaneously smooth analytic function of temperature what additionally excludes a possibility of a continuous phase transition. It could be thus concluded that the one-dimensional Ising spin systems are also free of any finite-temperature phase transition. This statement is in agreement with non-existence theorems for a phase transition of one-dimensional lattice-statistical models with short-range interactions due to Ruelle \cite{ruel68,ruel89}, Dyson \cite{dyso69}, Cuesta and Sanchez \cite{cues03}. Bearing this in mind, one should take with an extraordinary caution a substantial list of existing literature on one-dimensional Ising spin systems mostly referred to as nanotubes \cite{nt11a,nt11b,nt11c,nt11d,nt11e,nt12a,nt12b,nt12c,nt12d,nt13a,nt13b,nt13c,nt13d,nt13e,nt14a,nt14b,nt14c,nt15,nt16,nt17,nt18a,nt18b,nt19,nt20a,nt20b,nt21}, nanowires \cite{nw10a,nw10b,nw11,nw12a,nw12b,nw12c,nw12d,nw12e,nw12f,nw12g,nw12h,nw13a,nw13b,nw13c,nw13d,nw13e,nw14a,nw14b,nw14c,nw14d,nw14e,nw14f,nw14g,nw15a,nw15b,nw15c,nw15d,nw15e,nw15f,nw15g,nw16a,nw16b,nw16c,nw16d,nw16e,nw16f,nw16g,nw16h,nw16i,nw16j,nw16k,nw17a,nw17b,nw17c,nw17d,nw17e,nw17f,nw17g,nw17h,nw18a,nw18b,nw19,nw20}, nanoladders \cite{nl19a,nl19b,nl19c} or nanoribbons \cite{nr17,nr20}, for which the EFT erroneously predicts a spontaneous long-range order with nonzero critical temperature.

\section{Conclusion}
\label{sec:5}

In the present article we have investigated a few paradigmatic examples of zero- and one-dimensional Ising spin nanosystems with the help of EFT and rigorous calculation methods. More specifically, our exact calculations have convincingly evidenced non-existence of a spontaneous long-range order and a finite-temperature phase transition for zero-dimensional Ising spin systems such as star, cube, decorated hexagonal nanoparticle, star of David, as well as, one-dimensional Ising spin systems such as branched chain, sawtooth chain, two-leg ladder and hexagonal ladder. All considered examples of the Ising spin nanosystems thus clearly exemplify a failure of the EFT in predicting a spurious spontaneous long-range order and finite-temperature phase transition. 

The main purpose of the present work is not to criticize the EFT rather than to draw attention to its serious deficiency. The failure of the EFT closely relates to the fact that it predicts presence of a spontaneous long-range order and a finite-temperature phase transition for any Ising spin system involving at least one spin with the coordination number three regardless of its magnetic dimensionality. The false spontaneous long-range ordering and finite-temperature criticality is therefore predicted also for zero- or one-dimensional Ising spin systems. Bearing all this in mind, one should resort to more rigorous methods (exact enumeration, combinatorial, graph-theoretical, transfer-matrix methods, etc.) when treating zero- or one-dimensional Ising spin systems. 

In spite of this obvious failure, the EFT and its various extensions will still keep prominent position in statistical mechanics of the Ising spin systems, because their conceptual simplicity and higher precision compared to the traditional mean-field theory justify their applications to diverse more complex two- and three-dimensional Ising spin models. One should however avoid in future application of the EFT to zero- or one-dimensional Ising spin systems to get rid of a spurious spontaneous long-range order and finite-temperature phase transition. The comprehensive list of scientific literature, where the false conclusion about presence of spontaneous long-range order and finite-temperature phase transition was reached on grounds of the EFT, should be regarded as vehement warning and this clear mistake should not be widespread further. 

\section*{Acknowledgment}
This work was financially supported by the grant of The Ministry of Education, Science, Research and Sport of the Slovak Republic under the contract No. VEGA 1/0531/19 and by the grant of the Slovak Research and Development Agency under the contract No. APVV-16-0186.


\end{document}